\documentclass[fleqn,usenatbib]{mnras}

\usepackage{newtxmath}

\usepackage[T1]{fontenc}
\usepackage{ae,aecompl}

\usepackage{graphicx}	
\usepackage{amsmath}	
\usepackage{amssymb}	

\usepackage{natbib}
\usepackage{color}
\usepackage{rotating}
\usepackage{url}
\usepackage{ifthen}

\usepackage{bm}
\usepackage{bbold}
\usepackage{aas_macros}
\usepackage{verbatim}
\usepackage{algorithm}
\usepackage{algpseudocode}
\usepackage{enumitem}

\DeclareMathAlphabet{\mathpzc}{OT1}{pzc}{m}{it}

\newcommand\Wtilde{\stackrel{\sim}{\smash{\mathpzc{W}}\rule{0pt}{1.1ex}}}

\newcommand{\mvec}[1]{\bm{#1}}
\newcommand{\mmat}[1]{\mathbf{#1}}

\newcommand{\myvec}[1]{\mvec{#1}}
\newcommand{\mymat}[1]{\mmat{#1}}
\newcommand\numberthis{\addtocounter{equation}{1}\tag{\theequation}}

\SetSymbolFont{symbols}{bold}{OMS}{cmsy}{b}{n}
\DeclareSymbolFont{bmisymbols}{OML}{cmm}{b}{it}

\title[Optimal and fast $\mathcal{E}/\mathcal{B}$ separation]{Optimal and fast $\mathcal{E}/\mathcal{B}$ separation with a dual messenger field}

\author[D. Kodi Ramanah, G. Lavaux, B. D. Wandelt]{Doogesh Kodi Ramanah,$^{1,2}$\thanks{ramanah@iap.fr} Guilhem Lavaux,$^{1,2}$\thanks{lavaux@iap.fr} Benjamin D. Wandelt$^{1,2,3}$\\
$^{1}$ Sorbonne Universit\'e, CNRS, UMR 7095, Institut d'Astrophysique de Paris, 98 bis bd Arago, 75014 Paris, France\\
$^{2}$ Sorbonne Universit\'es, Institut  Lagrange  de  Paris  (ILP),  98  bis bd Arago, 75014 Paris, France\\
$^{3}$ Center for Computational Astrophysics, Flatiron Institute, 162 5th Avenue, 10010, New York, NY, USA\\
}

\date{Accepted XXX. Received YYY; in original form ZZZ}

\pubyear{2017}

\begin{document}
\label{firstpage}
\pagerange{\pageref{firstpage}--\pageref{lastpage}}
\maketitle

\begin{abstract}
We adapt our recently proposed dual messenger algorithm for spin field reconstruction and showcase its efficiency and effectiveness in Wiener filtering polarized cosmic microwave background (CMB) maps. Unlike conventional preconditioned conjugate gradient (PCG) solvers, our preconditioner-free technique can deal with high-resolution joint temperature and polarization maps with inhomogeneous noise distributions and arbitrary mask geometries with relative ease. Various convergence diagnostics illustrate the high quality of the dual messenger reconstruction. In contrast, the PCG implementation fails to converge to a reasonable solution for the specific problem considered. The implementation of the dual messenger method is straightforward and guarantees numerical stability and convergence. We show how the algorithm can be modified to generate fluctuation maps, which, combined with the Wiener filter solution, yield unbiased constrained signal realizations, consistent with observed data. This algorithm presents a pathway to exact global analyses of high-resolution and high-sensitivity CMB data for a statistically optimal separation of $\mathcal{E}$ and $\mathcal{B}$ modes. It is therefore relevant for current and next-generation CMB experiments, in the quest for the elusive primordial $\mathcal{B}$-mode signal. 
\end{abstract}




\section{Introduction}
\label{section1}

The polarization of the cosmic microwave background (CMB) radiation provides a window to probe the physics of the early Universe \citep[e.g.][]{hu1997primer, hu2002cmb, hu2003cmb} and consequently lies at the frontiers of research in modern cosmology. Polarization maps can provide more stringent constraints on cosmological parameters for cosmic variance limited experiments \citep[e.g.][]{galli2014cmb}, although their fascination lies primarily in the potential detection of the primordial $\mathcal{B}$-mode signals \citep[e.g.][]{guzzetti2016gravitational}. Such a measurement would potentially confirm the inflationary prediction of the imprint of the primordial gravitational waves on the CMB $\mathcal{B}$ modes on large angular scales \citep{kamionkowski2016quest}. The mathematical formalism of CMB polarization has been laid out around two decades ago by \cite{kamionkowski1997statistics} and \cite{zaldarriaga1997allsky}, with the link to primordial gravitational waves extensively investigated \citep[e.g.][]{kamionkowski1997probe, seljak1997signature}. 

Cosmological inference from current and next-generation CMB experiments tailored for mapping the polarized sky therefore requires sophisticated tools to optimize the scientific returns. An underlying issue is to extract the gradient and curl components, or the $\mathcal{E}$ and $\mathcal{B}$ modes, of the polarization signal from the data. A potential solution is provided by the Wiener filter \citep{wiener1949extrapolation}, a powerful signal reconstruction tool that incorporates statistical information about the signal and noise properties. It has widespread applications in cosmology and astrophysics, especially in the post-processing of observational data. 

Some common applications in the analysis of CMB data include map-making \citep[e.g.][]{bunn1994wiener, tegmark1997mapmaking}, optimal power spectrum estimation \citep[e.g.][]{tegmark1997power, bond1998estimating, oh1999efficient, elsner2012fastcalculation}, likelihood analysis \citep[e.g.][]{hinshaw2007threeyear, dunkley2009fiveyear, elsner2012likelihood}, treatment of foregrounds \citep[e.g.][]{bouchet1999multifrequency}, reconstruction of lensing potential, de-lensing and template-matching \citep[e.g.][]{hirata2003reconstruction, hirata2004cross, seljak2004gravitational, hanson2013detection, manzotti2017cmb, millea2017bayesian} and investigation of primordial non-Gaussianity \citep[e.g.][]{komatsu2005measuring, elsner2010improved, elsner2010local}. Moreover, we encounter the Wiener filter in Bayesian inference analyses involving the large-scale structures or the CMB \cite[e.g.][]{wandelt2004global, eriksen2004power, odwyer2004bayesian, jewell2004application, jasche2010bayesian, jasche2013methods, jasche2015matrix, anderes2015bayesian, alsing2016hierarchical, jasche2017foreground}. It has also been employed for the reconstruction of $21$-cm signal from contaminated data \citep[][]{gleser2008decontamination} and velocity field reconstructions \citep{lavaux2016virbius}. 

If we assume the observed data $\myvec{d}$ to be a linear combination of the signal $\myvec{s}$ and noise $\myvec{n}$, i.e.
\begin{equation}
	\myvec{d} = \myvec{s} + \myvec{n}, 
	\label{eq:data_model_DMPol}
\end{equation} 
then the Wiener filter equation can be written as follows:
\begin{equation}
	(\mymat{S}^{-1} + \mymat{N}^{-1})\myvec{s}_{\text{\tiny {\textup{WF}}}} = \mymat{N}^{-1}\myvec{d}. 
	\label{eq:wf_equation_DMPol}
\end{equation} 
The Wiener filter solution, $\myvec{s}_{\text{\tiny {\textup{WF}}}}$, is the maximum {\it a posteriori} solution in a Bayesian analysis where the signal and noise are both Gaussian random fields, with corresponding covariances $\mymat{S}$ and $\mymat{N}$. $\myvec{s}_{\text{\tiny {\textup{WF}}}}$ therefore maximizes the posterior probability distribution $\propto \exp(-\chi^2 /2)$, or equivalently minimizes:
\begin{equation}
	{\chi^2_{\tiny {}}} = (\myvec{d} - \myvec{s})^{\dagger}\mymat{N}^{-1}  (\myvec{d} - \myvec{s}) + \myvec{s}^{\dagger}\mymat{S}^{-1} \myvec{s}.
	\label{eq:chi2_WF_DMPol}
\end{equation}
$\myvec{s}_{\text{\tiny {\textup{WF}}}}$ is the least-square optimal solution, as no other linear solution has reconstruction errors with smaller mean-square deviations. 

The numerical issues encountered in computing the exact Wiener filter solution are well-documented in \cite{EW12} and \cite{DKR2017reloaded}. Traditional methods involve the use of the preconditioned conjugate gradient (PCG) scheme \citep[e.g.][]{eriksen2004power, wandelt2004global} and variants thereof, such as multi-scale \citep{smith2007background} and multi-grid algorithms \citep{seljebotn2014multi}. A pseudo-inverse based preconditioner has recently been implemented by \cite{seljebotn2017multi} for CMB component separation. The complexity of this method is illustrated by the variety of preconditioners that exist in the literature.   

The messenger method, first proposed by \cite{EW12}, is a preconditioner-free Wiener filtering technique. In a recent work, we presented the dual messenger algorithm, an enhanced variant of the standard messenger approach, as a general-purpose tool for Wiener filtering with applications in various areas of astrophysics and cosmology, with the focus being on the formalism and convergence properties of the algorithm \citep{DKR2017reloaded} (hereafter KLW17). We demonstrated the efficiency, effectiveness and unconditional stability of the dual messenger scheme with respect to the PCG approach when analysing CMB temperature maps. 

The extension of the dual messenger algorithm for the analysis of polarized CMB data sets is the next key step. Moreover, \cite{bunn2016pure} have recently shown that the application of the Wiener filter to CMB polarization data may produce pure $\mathcal{E}$ and $\mathcal{B}$ maps, free from cross-contamination, at a much lower computational cost than the standard methods \cite[e.g.][]{lewis2002analysis, lewis2003harmonic, bunn2003pixelised, bunn2011efficient}. This provides further motivation for an efficient polarized Wiener filtering tool. 

The inclusion of polarization data significantly increases the condition number of the covariance matrices involved \citep{larson2007estimation} such that finding appropriate preconditioners becomes highly non-trivial \citep{oh1999efficient}. As a result, the PCG method may require expensive pre-computations \citep[e.g.][]{seljebotn2014multi} and may be prone to numerical instabilities, as also illustrated in this work. The messenger techniques circumvent this ill-conditioning predicament with relative ease since they do not require any preconditioning, as demonstrated by \cite{EW12}. They implemented the standard messenger method for the Wiener filtering of polarized maps from WMAP, but they excluded $\mathcal{B}$ modes from the analysis, thereby reducing the dimension of the signal covariance.

Messenger techniques are becoming increasingly popular and are being further developed as a viable solution to complex and realistic problems. Recently, \cite{huffenberger2017cosmic} adapted the standard messenger method for map-making applications. Using mock Advanced ACTPol data, they illustrated the superior quality of maps obtained relative to a traditional PCG approach, thereby showcasing messenger field map-making as a potentially powerful CMB data analysis tool. \cite{huffenberger2017preconditionerfree} also applied the messenger method to problems with multiple, uncorrelated noise sources. 

In this work, we adapt our recently proposed dual messenger algorithm for spin field reconstruction. We consider an artificially generated polarized CMB data set with correlated noise and distinct temperature and polarization masks, while incorporating the $\mathcal{B}$ modes in the polarization signal. We demonstrate the ease of implementation, efficiency and unconditional stability of the algorithm. As a comparison, we also implement a PCG method and illustrate the difficulties encountered in converging to a plausible solution.

The paper is structured as follows. In Section \ref{section2}, we provide a brief description of the dual messenger algorithm and outline its numerical implementation for polarized signal reconstruction and generation of constrained realizations. We then showcase the capabilities of our scheme in Section \ref{section3}. Finally, we summarize our main findings in Section \ref{section4}. In Appendix \ref{jacobi_appendix}, we describe how the solution can be further refined via a modified Jacobi relaxation scheme. We also describe a generalized procedure for dealing with masks in Appendix \ref{mask_formalism_appendix} and provide the preconditioner adopted for the PCG implementation in Appendix \ref{preconditioner_appendix}.

\section{Dual Messenger Algorithm}
\label{section2}

The essence of the messenger methods lies in the introduction of an auxiliary field that acts as a mediator between the different bases where the signal and noise covariances, $\mymat{S}$ and $\mymat{N}$, can be conveniently expressed as sparse matrices. This essentially splits the Wiener filter equation into a set of algebraic equations that must be solved iteratively, obviating the need for matrix inversions or preconditioners. 

As pointed out in KLW17, the formalism of the dual messenger algorithm remains invariant with the inclusion of polarization data in the analysis. We briefly review the general system of equations that yields the two key equations to be implemented. A complementary and more in-depth description of the dual messenger algorithm is provided in KLW17. With the introduction of the messenger fields $\myvec{t}$ and $\myvec{u}$ at the level of the noise and signal, respectively, the modified $\chi^2$ is as follows:
\begin{multline}
	{\chi^2_{\tiny {T, U}}} = (\myvec{d} - \myvec{t})^{\dagger}\bar{\mymat{N}}^{-1}  (\myvec{d} - \myvec{t}) + (\myvec{t} - \myvec{u})^{\dagger} {\mymat{T}}^{-1}(\myvec{t} - \myvec{u}) \\ + (\myvec{u} - \myvec{s})^{\dagger}\mymat{U}^{-1} (\myvec{u} - \myvec{s}) + \myvec{s}^{\dagger} \bar{\mymat{S}}^{-1} \myvec{s},
	\label{eq:chi2_hybrid_messenger_DMPol}
\end{multline}
where we defined $\bar{\mymat{N}} \equiv \mymat{N} - \mymat{T}$, and we choose the covariance matrix of the auxiliary field $\myvec{t}$ according to $\mymat{T} = \alpha \mathbb{1}$, where $\alpha \equiv \textup{min}(\textup{diag}(\mymat{N}))$, and in analogous fashion, $\mymat{U} = \nu\mathbb{1}$ with $\nu \equiv \mathrm{min}(\textup{diag}(\mymat{S}))$ for the second auxiliary field $\myvec{u}$. The covariance of the signal $\myvec{s}$ is then given by $\bar{\mymat{S}} \equiv \mymat{S} - \mymat{U}$. Physically, $\myvec{t}$ corresponds to a homogeneous component of the noise covariance, while $\myvec{u}$ is the analogous component associated with the signal covariance. 

Minimizing the $\chi^2$ with respect to $\myvec{s}$, $\myvec{t}$ and $\myvec{u}$ leads to the following three equations:
\begin{align}
	\myvec{s} &= (\mymat{U}^{-1} + \bar{\mymat{S}}^{-1})^{-1} \mymat{U}^{-1} \myvec{u} \label{eq:1st_equation_hybrid_messenger_DMPol}\\
	\myvec{u} &= (\mymat{U}^{-1} + \mymat{T}^{-1})^{-1} ( \mymat{T}^{-1} \myvec{t} + \mymat{U}^{-1} \myvec{s} ) \label{eq:2nd_equation_hybrid_messenger_DMPol}\\
	\myvec{t} &= (\bar{\mymat{N}}^{-1} + \mymat{T}^{-1})^{-1} ( \mymat{T}^{-1} \myvec{u} + \bar{\mymat{N}}^{-1} \myvec{d} ) \label{eq:3rd_equation_hybrid_messenger_DMPol}.
\end{align}

We can reduce the above set of three equations to two equations, as in KLW17, as we need only one messenger field, but here, we contract the equations in an alternative way, to provide a more numerically convenient form of the equations adapted to deal with ill-conditioned systems, typical of CMB polarization problems, while improving convergence. By plugging equation (\ref{eq:1st_equation_hybrid_messenger_DMPol}) in equation (\ref{eq:2nd_equation_hybrid_messenger_DMPol}) and defining $\bm{\xi} = \mymat{U} + \mymat{T}$, we obtain the following set of two equations to be solved iteratively:
\begin{align}
	\myvec{u} &= (\bar{\mymat{S}} + \mymat{U}) (\bar{\mymat{S}} + \bm{\xi})^{-1} \myvec{t} \label{eq:reduced_hybrid_messenger_1st_equation_DMPol}\\
	\myvec{t} &= ( {\bar{\mymat{N}}}^{-1} + \mymat{T}^{-1} )^{-1} (\mymat{T}^{-1} \myvec{u} + {\bar{\mymat{N}}}^{-1} \myvec{d}) ,
	\label{eq:reduced_hybrid_messenger_2nd_equation_DMPol}
\end{align}
where equation (\ref{eq:reduced_hybrid_messenger_1st_equation_DMPol}) is simply a Wiener filter of $\myvec{t}$, assuming a modified signal covariance $( \bar{\mymat{S}} + \mymat{U} )$. The mechanism adopted to improve convergence is as follows: We artificially truncate the signal covariance $\mymat{S}$ to some lower initial value of $\ell_\mathrm{iter}$ that corresponds to a covariance $\mu$. By implementing a cooling scheme for $\bm{\xi}$, we subsequently vary $\mymat{U}$ to bring $\mu \rightarrow \nu$, where, in the limit $\mu = \nu$, we have $\myvec{u} = \myvec{s}$ and the above system of equations (\ref{eq:reduced_hybrid_messenger_1st_equation_DMPol}) and (\ref{eq:reduced_hybrid_messenger_2nd_equation_DMPol}) reduces to the usual Wiener filter equation (\ref{eq:wf_equation_DMPol}). This results in a redefinition of $\bar{\mymat{S}}$ using the Heaviside function as $\bar{\mymat{S}} = \Theta (\mathcal{S} - \mymat{U})$, where $\mathcal{S}$ corresponds to the eigenvalues of $\mymat{S}$. This leads to a hierarchical framework, where we obtain the solution on the largest scales initially, and gradually the algorithm resolves the fine structures on the small scales. 

Due to the continuous mode of the signal, i.e. the zero eigenvalue of $\mymat{S}$, we therefore require $\mu \rightarrow \nu = 0$ to finally obtain the desired Wiener filter solution. The cooling scheme for $\bm{\xi}$ involves reducing $\bm{\xi}$ by a constant factor and iterating until $\bm{\xi} \rightarrow \mymat{T}$, at which point $\mu = 0$, as desired. Moreover, $\mymat{U}$ does not need to be strictly proportional to the identity matrix, and this useful property allows us to solve the temperature and polarization signals at different rates. The rationale behind the above approach is described quantitatively in KLW17.

Note that the above equations, for simplicity, are written in a single basis, but all the operators are written in their respective bases, with all basis transformations made explicit, in Algorithm \ref{alg:dual_messenger_DMPol} below. In terms of the numerical implementation for the joint temperature and polarization analysis, the formalism of the signal and data vectors, $\myvec{s}$ and $\myvec{d}$, signal and noise covariances, $\mymat{S}$ and $\mymat{N}$, must be generalized as described below. 

\subsection{Numerical implementation}
\label{section2.1}

The CMB signal can be described as a $3 N_{\mathrm{pix}}$ dimensional vector of harmonic coefficients, $\myvec{s}_{\ell} = ( a_{\ell}^{\mathcal{T}}, a_{\ell}^{\mathcal{E}}, a_{\ell}^{\mathcal{B}} )$, for each $\ell$, where $\mathcal{T}$, $\mathcal{E}$ and $\mathcal{B}$ imply temperature, electric/gradient, and magnetic/curl, respectively, for discretized sky maps of $N_{\mathrm{pix}}$ pixels. The data, $\myvec{d} = ( \myvec{d}_{\mathcal{I}}, \myvec{d}_{\mathcal{Q}}, \myvec{d}_{\mathcal{U}} )$, are pixelized maps of the Stokes parameters, $\mathcal{I}$, $\mathcal{Q}$ and $\mathcal{U}$, where $\mathcal{I}$ here corresponds to the temperature anisotropy.  

Since we assume that the CMB anisotropies are isotropic and Gaussian, the signal covariance $\mymat{S}$ is diagonal in Fourier space, in the flat-sky approximation. $\mymat{S}$ becomes a block-diagonal matrix, with a $3 \times 3$ sub-matrix,  for all multipole moments $\ell$, as follows:
\begin{equation}
	\mymat{S}_{\ell} = \begin{pmatrix}
C_{\ell}^{TT} & C_{\ell}^{TE} & 0 \\ 
C_{\ell}^{TE} & C_{\ell}^{EE} & 0 \\ 
0 & 0 & C_{\ell}^{BB} 
\end{pmatrix}, 
	\label{eq:signal_covariance_DMPol}
\end{equation} 
with the vanishing cross-spectra, $C_{\ell}^{TB}$ and $C_{\ell}^{EB}$, set to zero.

Here, we consider correlated noise, where the noise covariance matrix is a $3 \times 3$ block-diagonal matrix, with non-zero elements. The noise covariance matrix has the following block-diagonal structure, for every pixel $i$:
\begin{equation}
	\mymat{N}_{i} = \begin{pmatrix}
\left\langle \mathcal{II} \right\rangle & \left\langle \mathcal{IQ} \right\rangle & \left\langle \mathcal{IU} \right\rangle \\ 
\left\langle \mathcal{QI} \right\rangle & \left\langle \mathcal{QQ} \right\rangle & \left\langle \mathcal{QU} \right\rangle \\ 
\left\langle \mathcal{UI} \right\rangle & \left\langle \mathcal{UQ} \right\rangle & \left\langle \mathcal{UU} \right\rangle 
\end{pmatrix}.
	\label{eq:noise_covariance_DMPol}
\end{equation} 

The numerical implementation of the dual messenger algorithm adapted for polarized signal reconstruction is outlined in Algorithm \ref{alg:dual_messenger_DMPol}. In particular, the computations of $\bar{\mymat{S}}$ and $\bar{\mymat{N}}$, via their respective diagonalized forms $\bar{\mathcal{S}}$ and $\bar{\mathcal{N}}$, are clearly illustrated. Since the algorithm requires $\bar{\mathcal{N}}^{-1}$, we circumvent the corner case resulting from $\bar{\mathcal{N}} = \bm{0}$, i.e. when $\mathcal{N} = \mymat{T}$, by imposing the following constraint on $\myvec{t}$ in the correct vector subspace: $\mathfrak{D} \myvec{t} |_{\mathcal{N} = \mymat{T}} = \mathfrak{D} \myvec{d} |_{\mathcal{N} = \mymat{T}}$, following the notation set in Algorithm \ref{alg:dual_messenger_DMPol}. 

In Appendix \ref{jacobi_appendix}, we provide a brief description of how the algorithm can be embedded in a modified Jacobi relaxation mechanism to further refine the solution. We implemented the mask by setting the noise covariance for the masked pixels to a numerically high value $\sim \mathcal{O}(10^{10})$. While the formalism above, as illustrated in Algorithm \ref{alg:dual_messenger_DMPol}, is still valid for dealing with a common temperature and polarization mask, where the inverse noise covariance must be set to zero, it is nevertheless not completely adequate for dealing with different masks. The procedure for solving the messenger equation (\ref{eq:reduced_hybrid_messenger_2nd_equation_DMPol}) must be consequently modified as described in Appendix \ref{mask_formalism_appendix}. We will implement this exact masking procedure in a forthcoming publication where complex noise covariances will be considered for the analysis of real data sets. 

A brief note concerning Fourier transforms: Defining $\phi_{\ell}$ as the angle between the vector $\myvec{\ell}$ and the $\ell_x$ axis, we use the following convention for the Fourier transforms of the corresponding maps of the Stokes parameters:
\begin{align}
	\widehat{\mathcal{I}}_{\myvec{\ell}} &= \mathpzc{F}^{\dagger} \mathcal{I}_{\myvec{x}} = \left( \frac{1}{L^2} \right) \sum_{\myvec{x}} \omega^{-\myvec{x}\cdot\myvec{k}} \mathcal{I}_{\myvec{x}} \label{eq:fourier_I_DMPol}\\
	\widehat{\mathcal{Q}}_{\myvec{\ell}} &= \mathpzc{F}^{\dagger} \mathcal{Q}_{\myvec{x}} = \left( \frac{1}{L^2} \right) \sum_{\myvec{x}} \left( a_{\myvec{\ell}}^{\mathcal{E}} \cos 2\phi_{\ell} - a_{\myvec{\ell}}^{\mathcal{B}} \sin 2\phi_{\ell} \right) \omega^{-\myvec{x}\cdot\myvec{k}} \label{eq:fourier_Q_DMPol}\\
	\widehat{\mathcal{U}}_{\myvec{\ell}} &= \mathpzc{F}^{\dagger} \mathcal{U}_{\myvec{x}} = \left( \frac{1}{L^2} \right) \sum_{\myvec{x}} \left( a_{\myvec{\ell}}^{\mathcal{E}} \sin 2\phi_{\ell} + a_{\myvec{\ell}}^{\mathcal{B}} \cos 2\phi_{\ell} \right) \omega^{-\myvec{x}\cdot\myvec{k}} \label{eq:fourier_U_DMPol}
\end{align}
for $\omega = \exp({i2\pi / N_{\mathrm{pix}}})$ and an observed sky patch of angular extent $L$, such that, for the harmonic coefficients, we have:
\begin{align}
	a_{\myvec{\ell}}^{\mathcal{E}} &= \widehat{\mathcal{Q}}_{\myvec{\ell}} \cos 2\phi_{\ell} + \widehat{\mathcal{U}}_{\myvec{\ell}} \sin 2\phi_{\ell} \label{eq:fourier_E_DMPol}\\
	a_{\myvec{\ell}}^{\mathcal{B}} &= - \widehat{\mathcal{Q}}_{\myvec{\ell}} \sin 2\phi_{\ell} + \widehat{\mathcal{U}}_{\myvec{\ell}} \cos 2\phi_{\ell} , \label{eq:fourier_B_DMPol}
\end{align}
and trivially, $a_{\myvec{\ell}}^{\mathcal{T}} = \widehat{\mathcal{I}}_{\myvec{\ell}}$. The corresponding inverse Fourier transform, $\mathpzc{F}$, satisfies $\mathpzc{F}\mathpzc{F}^{\dagger} = \sigma \mathbb{1}$, where $\sigma = N_{\mathrm{pix}}^2/L^4$. From a linear algebraic standpoint, we have two operators $\mathcal{F}^{\dagger}$ and $\mathcal{F}$ that act on the vector spaces $(\mathcal{I},\mathcal{Q},\mathcal{U})$ and $(a_{\ell}^{\mathcal{T}}, a_{\ell}^{\mathcal{E}}, a_{\ell}^{\mathcal{B}})$, respectively, while still satisfying the orthogonality condition, $\mathcal{F}\mathcal{F}^{\dagger} = \sigma \mathbb{1}$.   

\begin{algorithm}
\begin{algorithmic}[1]
  \Procedure{dual messenger}{$\myvec{d}$, $\mymat{N}$, $\mymat{S}, N_{\mathrm{pix}}, L$}
  \State $\myvec{s}_0 = \mathrm{zeros}(N_{\mathrm{pix}},N_{\mathrm{pix}},3)$ \Comment{Initialize $\myvec{s}$ with zeros}
  \State $\myvec{t}_0 = \myvec{d}$ \Comment{Initialize $\myvec{t}$ via an initial guess}
  \State \Comment{Diagonalize $\mymat{N}$ via basis transformation}
  \State $\mymat{N} = \mathfrak{D}^{\dagger} \mathcal{N} \mathfrak{D}$ 
  \State \Comment{Compute the covariance of messenger field $\myvec{t}$}
  \State $\alpha = \mathrm{min}(\mathcal{N})$ \Comment{such that $\mymat{T} = \alpha \mathbb{1}$} 
  \State $\bar{\mathcal{N}} = \mathcal{N} - \mymat{T}$ \Comment{Compute the covariance $\bar{\mathcal{N}}$}
  \State $\mymat{S} = \mathfrak{R}^{\dagger} \mathcal{S} \mathfrak{R}$ \Comment{Diagonalize $\mymat{S}$ via basis transformation}
  \State $\bm{\xi} = \bm{\mu} + \sigma^{-1} \alpha \mathbb{1} $ \Comment{Compute $\bm{\xi}$, $\bm{\mu} = ( \mu_{\mathcal{T}}, \mu_{\mathcal{E}}, \mu_{\mathcal{B}} )$}
  \State \Comment{$\sigma = N_{\mathrm{pix}}^2 / L^4$ is a numerical factor, $\mathcal{F}\mathcal{F}^{\dagger} = \sigma \mathbb{1}$}
  \While{$\bm{\xi}$ $\rightarrow$ $\sigma^{-1} \alpha \mathbb{1}$}
    \State $\mymat{U}_{\myvec{\ell}} = \bm{\mu}$ \Comment{Assign/update covariance $\mymat{U}$}     
    \State ${\bar{\mathcal{S}}}_{\myvec{\ell}} = \Theta (\mathcal{S} - \mymat{U})_{\myvec{\ell}}$ \Comment{Compute covariance $\bar{\mathcal{S}} \:$} 
  	\Repeat 
  	    \State \Comment{Transform to Fourier space} 
  	    \State \Comment {cf. equations (\ref{eq:fourier_I_DMPol}) - (\ref{eq:fourier_B_DMPol})} 
        \State $\hat{\myvec{s}}_{i+1,\myvec{\ell}} = \mathfrak{R}^{\dagger} \left(\bar{\mathcal{S}} + \mymat{U}\right)_{\myvec{\ell}} \left(\bar{\mathcal{S}} + \bm{\xi}\right)_{\myvec{\ell}}^{-1} \mathfrak{R} \left( \sigma^{-1} \mathcal{F}^{\dagger} \myvec{t}_{i,\myvec{x}} \right)_{\myvec{\ell}}$ 
        \State \Comment{Transform to pixel space}
       	\State $\myvec{s}_{i+1,\myvec{x}} = \mathcal{F}(\hat{\myvec{s}}_{i+1,\myvec{\ell}})$ 
       	\State $\myvec{t}_{i+1,\myvec{x}} = \mathfrak{D}^{\dagger} \left( \bar{\mathcal{N}}^{-1} + \mymat{T}^{-1} \right)_{\myvec{x}}^{-1}$
       	\State $\; \; \; \; \; \; \; \; \; \; \; \; \; \; \; \; \; \; \; \; \; \; \; \; \; \; \; \; \; \; \;  \cdot \left( \mathfrak{D} \mymat{T}^{-1}\myvec{s}_{i+1} + \bar{\mathcal{N}}^{-1} \mathfrak{D} \myvec{d} \right)_{\myvec{x}}$
       	\State $i \gets i+1$
    \Until{$\left\| \myvec{s}_{i} - \myvec{s}_{i-1} \right\| / \left\| \myvec{s}_{i} \right\| < \epsilon$}
    \State $\bm{\xi} \gets \bm{\xi} \times \beta$ \Comment{Cooling scheme for $\bm{\xi}$}
    \State $\bm{\mu} \gets \bm{\xi} - \sigma^{-1} \alpha \mathbb{1}$ \Comment{Compute resulting $\bm{\mu}$}
  \EndWhile 
      \State {$\myvec{s} \rightarrow \myvec{s}_{\text{\tiny {\textup{WF}}}}$} \Comment{as $\bm{\xi} \rightarrow \sigma^{-1} \alpha \mathbb{1}, \bm{\mu} \rightarrow \bm{\nu} = \bm{0}$}
    \State \Return{$\myvec{s}_{\text{\tiny {\textup{WF}}}}$}
  \EndProcedure
\end{algorithmic}
\caption{\label{alg:dual_messenger_DMPol} Dual messenger algorithm}
\end{algorithm}

\subsection{Constrained realizations}
\label{section2.2}

It is well-known that the application of a Wiener filter leads to a reduced signal covariance,
\begin{equation}
	\langle \myvec{s}_{\text{\tiny {\textup{WF}}}} \myvec{s}_{\text{\tiny {\textup{WF}}}}^{\dagger} \rangle = \mymat{S} ( \mymat{S} + \mymat{N} )^{-1} \mymat{S}, 
	\label{eq:reduced_signal_covariance_DMPol}
\end{equation} 
by suppressing the power related to the noise. We therefore need to add a fluctuation vector $\myvec{f}$ to the Wiener-filtered maps to obtain signals that are consistent with the observed data, i.e. having the correct covariance properties, according to 
\begin{equation}
	\langle \myvec{f} \myvec{f}^{\dagger} \rangle = ( {\mymat{S}}^{-1} + {\mymat{N}}^{-1} )^{-1}. 
	\label{eq:fluctuation_vector_DMPol}
\end{equation} 
The dual messenger algorithm presented above can be augmented to generate full-sky, noiseless maps by making some minor adjustments. We simulate a fake signal $\hat{\myvec{s}}$ with the (prior) signal covariance $\mymat{S}$ assumed for the Wiener filter, which is subsequently contaminated with noise with covariance $\mymat{N}$ to generate a fake data set $\hat{\myvec{d}}$. We obtain constrained realizations \citep[e.g.][]{hoffman1991constrained} via
\begin{align*}
	\myvec{s}_{\text{\tiny {\textup{CR}}}} &= \myvec{s}_{\text{\tiny {\textup{WF}}}} + \myvec{f} \\
	&= \Wtilde\myvec{d} + (\hat{\myvec{s}} - \Wtilde\hat{\myvec{d}}) \\
	&= \Wtilde( \myvec{d} - \hat{\myvec{d}} ) + \hat{\myvec{s}}, \numberthis
	\label{eq:constrained_realization_DMPol}
\end{align*}
using the Wiener-filtered map, $\myvec{s}_{\text{\tiny {\textup{WF}}}} = \Wtilde\myvec{d}$, and the fluctuation map, $\myvec{f} = \hat{\myvec{s}} - \Wtilde \hat{\myvec{d}}$. The input to the dual messenger algorithm is therefore $(\myvec{d} - \hat{\myvec{d}})$, such that only one execution of the algorithm is sufficient to provide a constrained realization.

Now, the algorithm yields a solution that is a random realization of a fluctuation map with the correct signal properties. For a single $\myvec{d}$, the Wiener filter samples are plausible signal realizations that optimally take into account the constraints on the signal from the data and are hence known as constrained realizations.

\section{Application to CMB polarization}
\label{section3}

\begin{figure}
	\centering
		{\includegraphics[scale=0.7]{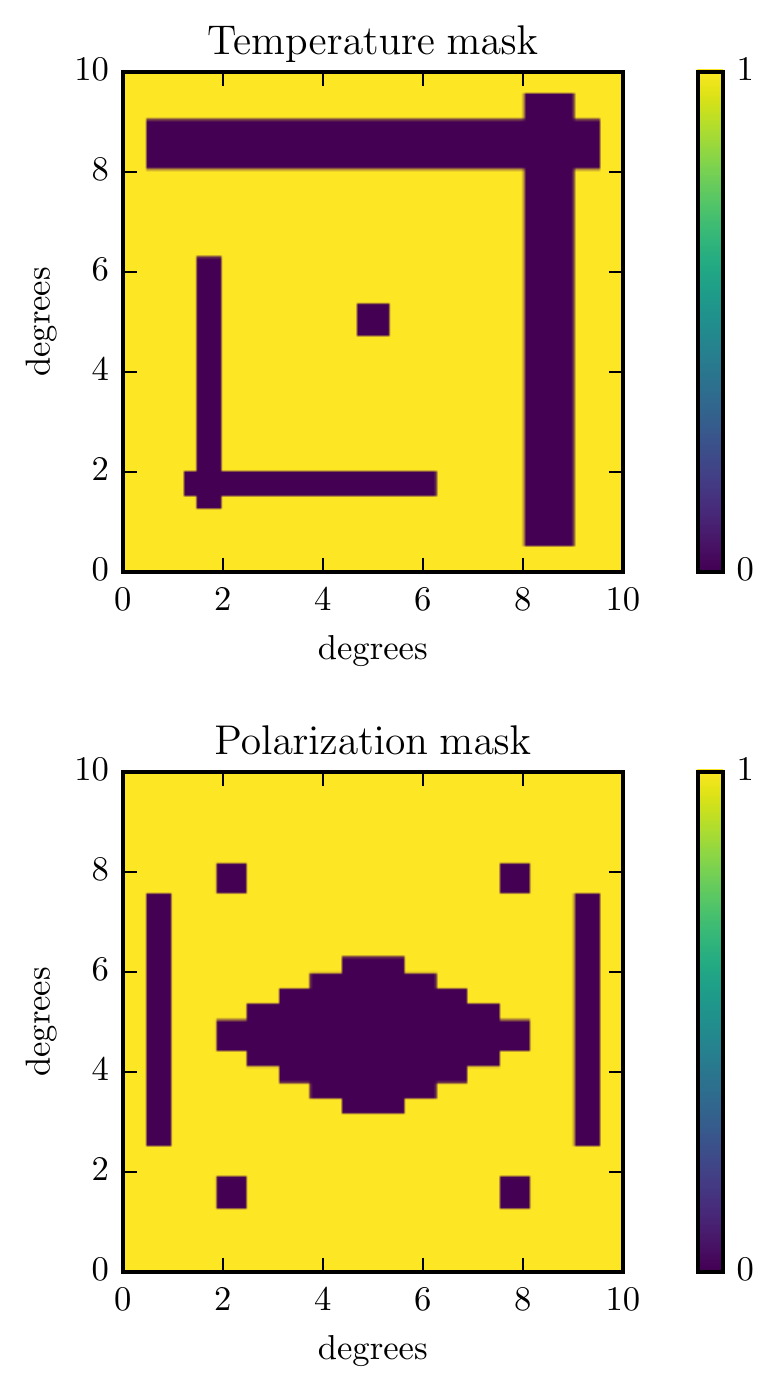}}
	\caption{The temperature and polarization masks implemented in the data analysis, corresponding to sky fractions of $f_{\mathrm{sky}}^{\mathcal{T}} = 0.78$ and $f_{\mathrm{sky}}^{\mathcal{P}} = 0.82$, respectively.}
	\label{fig:temp_pol_masks}
\end{figure}

\begin{figure*}
	\centering
		{\includegraphics[width=\hsize,clip=true]{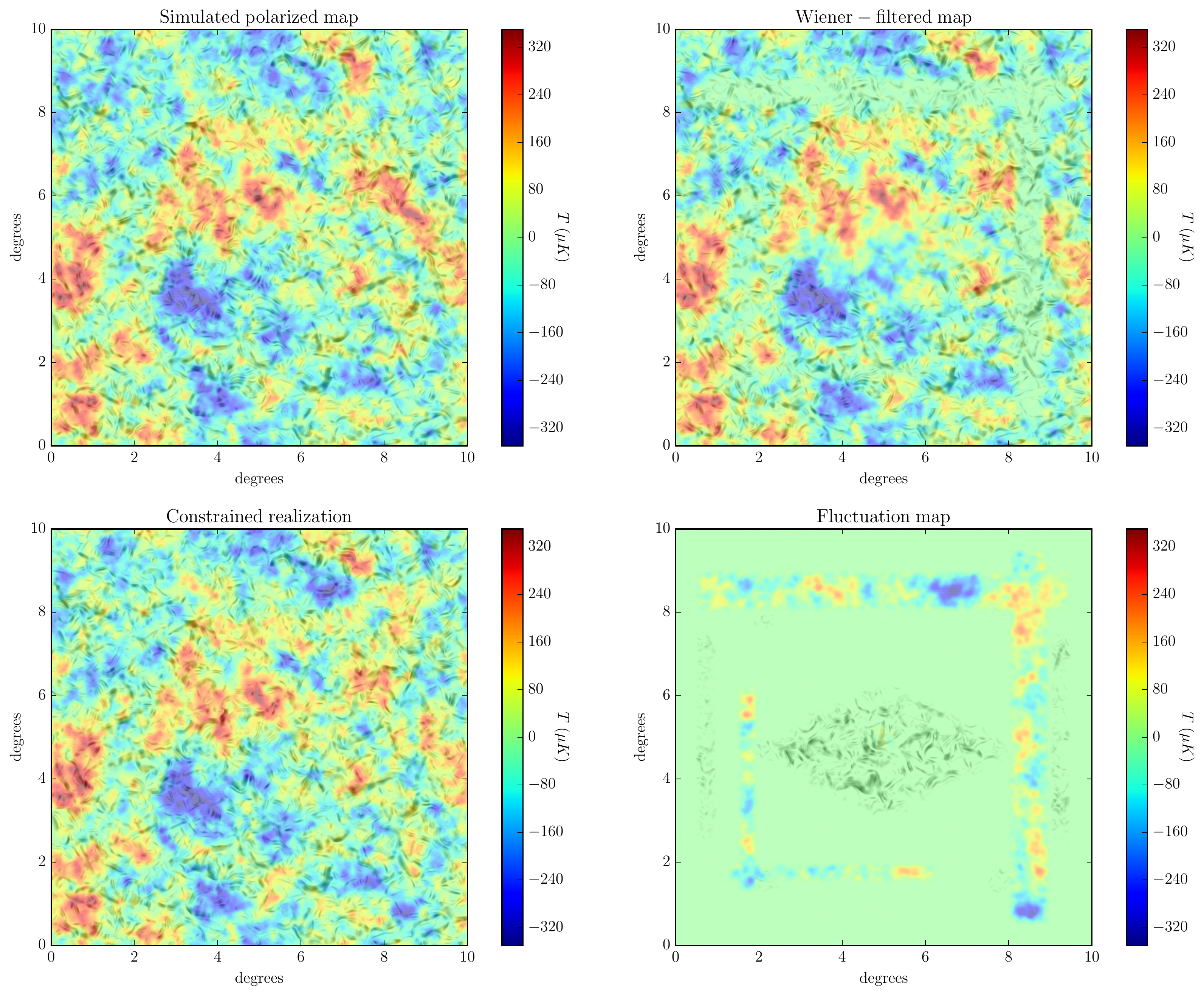}}
	\caption{{\it Top panel:} The simulated and Wiener-filtered polarization maps. The images above depict temperature as colour and polarization overlaid as a pattern of stripes. The alignment of the stripes indicates the direction of polarization while the level of transparency corresponds to the polarization intensity, with the darker regions implying stronger polarization. The Wiener-filtered map is the maximum {\it a posteriori} reconstruction of the signal from the data. {\it Bottom panel:} The fluctuation map and constrained realization. To compensate for the power loss due to noise and masked sky, the Wiener-filtered map is augmented with a fluctuation map, thereby yielding a full-sky, noiseless sample, i.e. constrained realization, with the correct signal properties. The temperature and polarization masks applied are visible in the fluctuation map.}
	\label{fig:pol_maps}
\end{figure*}

\begin{figure*}
	\centering
		{\includegraphics[width=\hsize,clip=true]{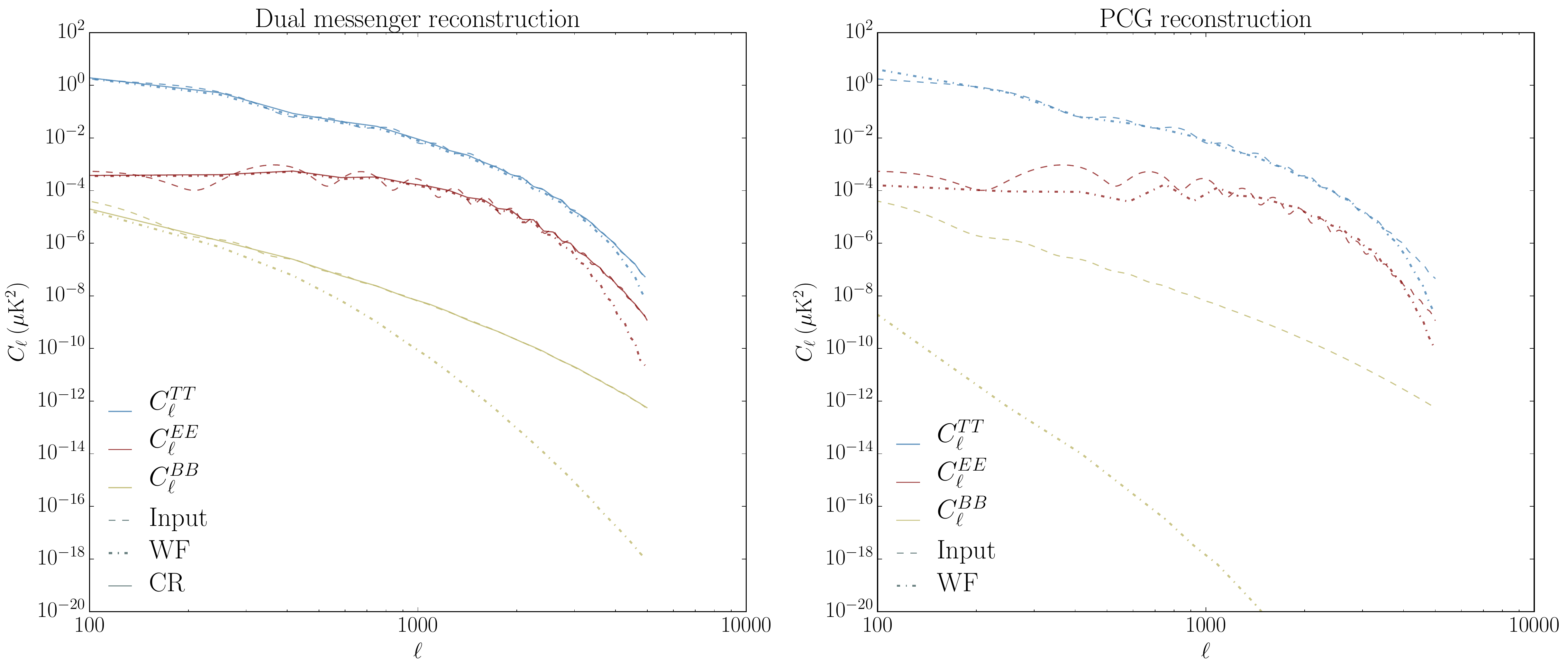}}
	\caption{Reconstructed (binned) power spectra computed using the dual messenger and PCG algorithms. The dashed lines indicate the input angular power spectra from which the polarized CMB signals are drawn, while the reconstructions from the Wiener-filtered (WF) and constrained realization (CR) maps are depicted by the corresponding dotted and solid lines. {\it Left panel:} We demonstrate that a constrained realization obtained via the dual messenger algorithm, as a combination of the Wiener-filtered and fluctuation maps, yields unbiased power spectra compared to the corresponding input power spectra. {\it Right panel:} The PCG reconstruction strikingly fails to find enough $\mathcal{B}$ modes, while the $\mathcal{E}$-mode power spectrum also displays some artefacts, especially on large and intermediate scales, demonstrating the unreliability of the power spectra reconstruction. Remarkably, the dual messenger reconstruction, even with a more lenient $\epsilon = 10^{-4}$, is visually undistinguishable from that displayed in the left panel with $\epsilon = 10^{-6}$, further highlighting the stark contrast in performance between these two methods.}
	\label{fig:C_l_recon}
\end{figure*}

\begin{figure*}
	\centering
		{\includegraphics[width=\hsize,clip=true]{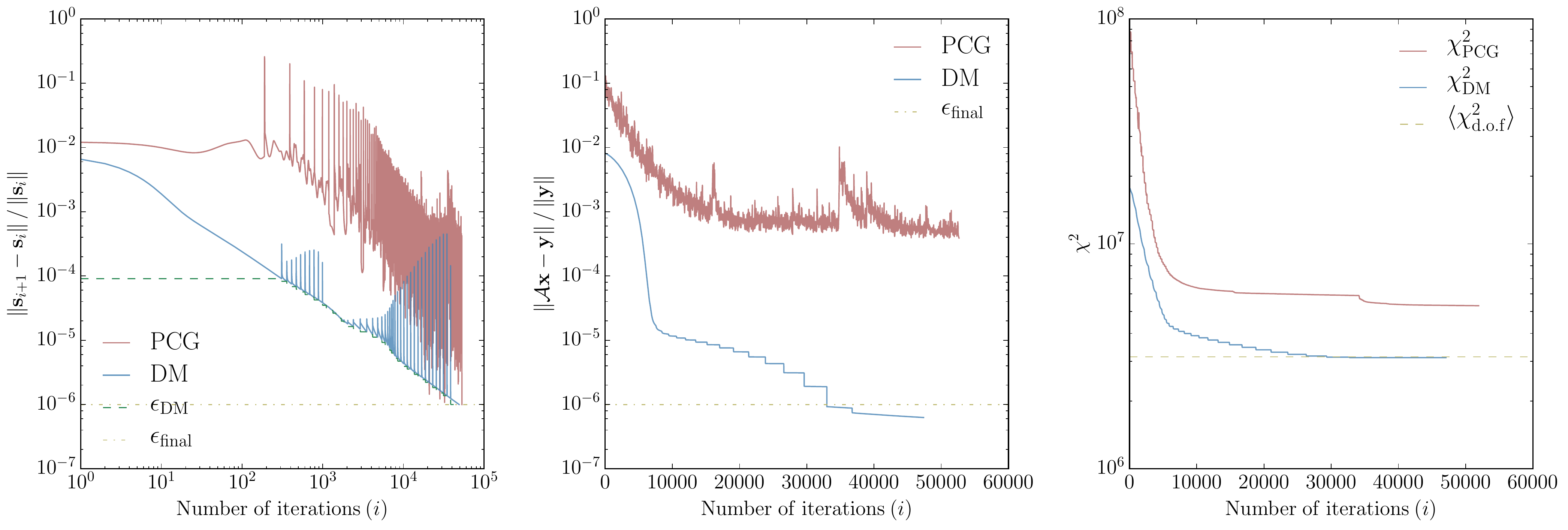}}
	\caption{Convergence diagnostics corresponding to the dual messenger (DM) and PCG reconstructions. {\it Left panel:} Variation of the residual error, given by the Cauchy criterion, with number of iterations. The cooling scheme for the convergence threshold $\epsilon$, indicated by the dashed lines, allows for faster convergence to the final tolerance desired. {\it Middle panel:} The corresponding variation of the residual error, given by $\left \| \mathcal{A} \myvec{x} - \myvec{y} \right\| / \left\| \myvec{y} \right\|$, with number of iterations. The monotonic decrease in this residual error demonstrates the unconditional stability of the dual messenger algorithm. In contrast, for the PCG counterpart, this residual error does not drop below $10^{-3}$, three orders of magnitude above that attained by the dual messenger solution, implying that the accuracy of the PCG solution is significantly lower. The PCG implementation is also susceptible to numerical instabilities, as indicated by the oscillations in the residual error. {\it Right panel:} Variation of $\chi^2$ with number of iterations. $\chi^2_{\mathrm{DM}}$ drops rapidly and finally matches the expectation value of the $\chi^2$ of the final solution, $\langle \chi^2_{\mathrm{d.o.f}} \rangle$, given by the number of degrees of freedom, indicated by the dashed line. $\chi^2_{\mathrm{PCG}}$, however, fails to achieve this expected level, again highlighting the unreliability and poor quality of the PCG solution.}
	\label{fig:convergence_diagnostics}
\end{figure*}

\begin{figure*}
	\centering
		{\includegraphics[width=\hsize,clip=true]{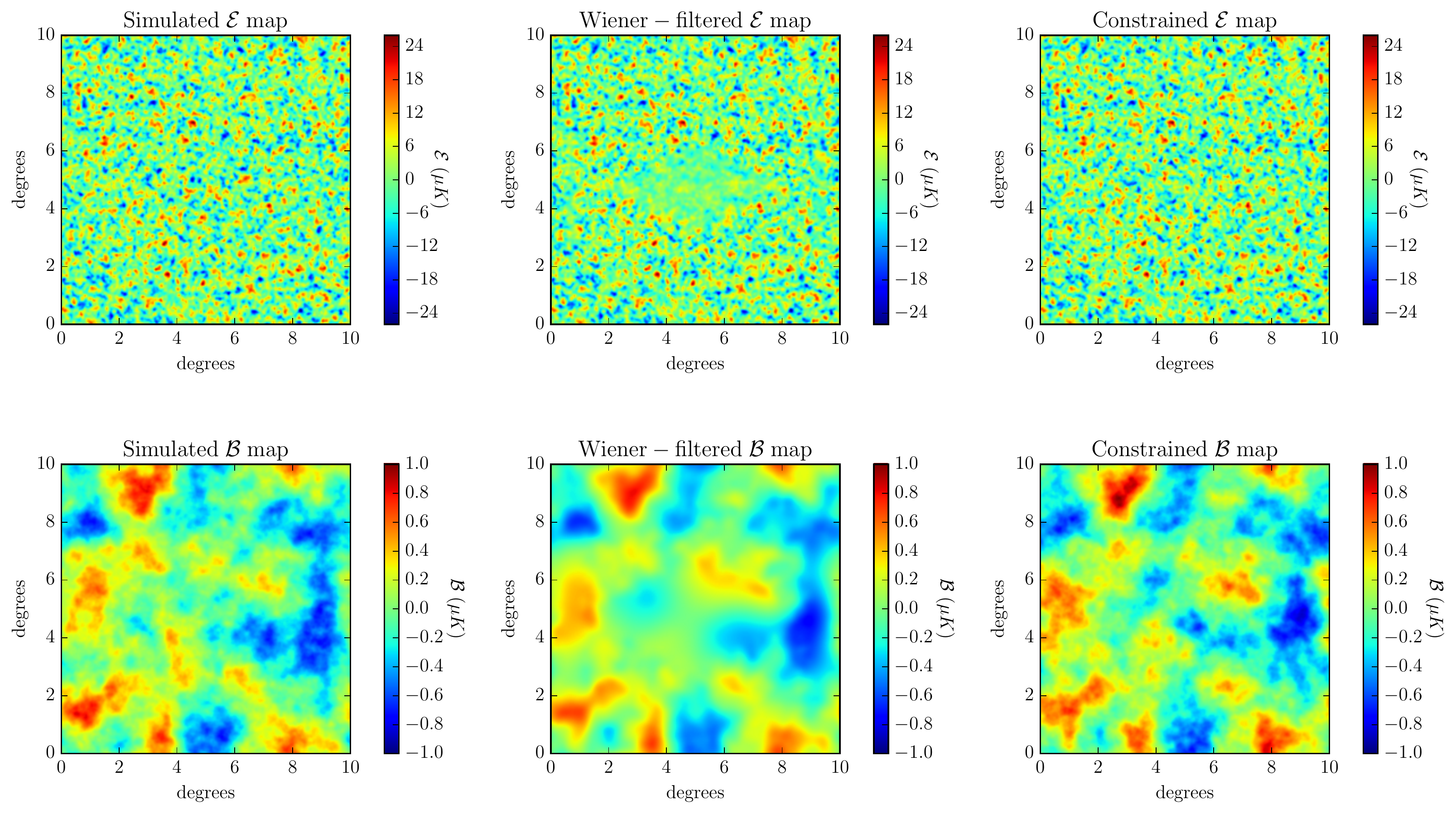}}
	\caption{The simulated, Wiener-filtered and constrained $\mathcal{E}$ and $\mathcal{B}$ maps. In the top row, the Wiener filter characteristically extends the signal into the masked regions; this works especially well in the small excized areas of the map, usually required for dealing with point foreground sources. Due to the non-vanishing cross spectrum $C_{\ell}^{TE}$, the polarization reconstruction in masked areas is rather efficient, unless the temperature and polarization masks overlap. In the bottom row, the low-amplitude $\mathcal{B}$ modes are inevitably smoothed out due to the suppression of the small-scale power. The constrained maps represent plausible realizations of de-noised full-sky $\mathcal{E}$ and $\mathcal{B}$ maps.}
	\label{fig:EB_maps}
\end{figure*}

\subsection{Map simulation}
\label{section3.1}

To simulate joint temperature and polarization maps in the flat-sky approximation, we make use of a Cholesky decomposition to generate realizations of $a_{\myvec{\ell}}^{\mathcal{T}}$, $a_{\myvec{\ell}}^{\mathcal{E}}$ and $a_{\myvec{\ell}}^{\mathcal{B}}$ signals with the correct covariance properties (cf. equation (\ref{eq:signal_covariance_DMPol})), taking into account the correlation between CMB temperature anisotropy and polarization. We made use of \textsc{camb}\footnote{http://camb.info} \citep{camb1999} to generate the input angular power spectra, $C^{TT}_{\ell}$, $C^{EE}_{\ell}$, $C^{BB}_{\ell}$ and $C^{TE}_{\ell}$, from which the corresponding CMB signals are drawn. We assume a standard $\Lambda$CDM cosmology with the set of cosmological parameters ($\Omega_{\mathrm{m}} = 0.32$, $\Omega_\Lambda = 0.69$, $\Omega_{\mathrm{b}} = 0.05$, $h = 0.67$, $\sigma_8 = 0.83$, $n_{\mathrm{s}} = 0.97$) from {\it Planck} \citep{13planck2015}. We can then construct the input $\mathcal{Q}$ and $\mathcal{U}$ maps by transforming realizations of $\mathcal{E}$ and $\mathcal{B}$ signals \citep[cf. Section IIc in][]{bunn2003pixelised} over flat-sky patches with angular extent, $L = 10.0$ degrees, and grid resolution, $N_{\textup{pix}} = 1024^2$. The input Stokes parameters' maps are subsequently contaminated with correlated noise, according to the noise covariance given by equation (\ref{eq:noise_covariance_DMPol}), with a noise amplitude of $4.0 \: \mu \mathrm{K}$, typical of high-sensitivity CMB experiments tailored for the detection of $\mathcal{B}$ modes. The temperature and polarization masks implemented, corresponding to sky fractions of $f_{\mathrm{sky}}^{\mathcal{T}} = 0.78$ and $f_{\mathrm{sky}}^{\mathcal{P}} = 0.82$, respectively, are depicted in Fig. \ref{fig:temp_pol_masks}.

\subsection{Polarization analysis}
\label{section3.2}

We showcase the application of the dual messenger algorithm in polarization data analysis, while drawing a comparison to the corresponding solution provided by a PCG method. For the PCG computation, we make use of the preconditioner provided in Appendix \ref{preconditioner_appendix}. 

As per standard data analysis pipelines, we must filter out the noise and reconstruct a clean  map via a Wiener filtering algorithm. We implement the dual messenger scheme described in the above sections to compute the Wiener-filtered $\mathcal{I}$, $\mathcal{Q}$ and $\mathcal{U}$ maps of the Stokes parameters. The algorithm loops through the iterations until the fractional difference between successive iterations has reached a sufficiently low value. Here, we implement Cauchy's ``weak'' criterion for convergence, $\left \| \myvec{s}_{i+1} - \myvec{s}_{i} \right\| / \left\| \myvec{s}_{i} \right\| < \epsilon$, where $\epsilon = 10^{-6}$. We adopt the same cooling scheme as in KLW17, whereby we reduce $\bm{\xi}$ by a constant factor, i.e. $\bm{\xi} \rightarrow \bm{\xi} \beta$, where $\beta = 3/4$, until $\bm{\xi} \rightarrow \sigma^{-1} \alpha \mathbb{1}$. Since we are not solving the desired system of equations initially, we can also implement a cooling scheme for the threshold $\epsilon$. This speeds up the computation significantly by around a factor of three. We relax the convergence criterion by a factor of $\eta$ for each $\mu$, thereby reducing $\epsilon$ from $10^{-4}$ to $10^{-6}$, where we choose $\eta = 1.1$.

The simulated and corresponding Wiener-filtered maps are displayed in the top row of Fig. \ref{fig:pol_maps}. The polarization intensity is given by $( \mathcal{Q}^2 + \mathcal{U}^2 )^{1/2}$, while the direction of polarization corresponds to $\arctan(\mathcal{Q} / \mathcal{U})/2 $. The Wiener-filtered map, as the maximum {\it a posteriori} reconstruction, represents the CMB signal content of the data, with the reconstruction of the large-scale modes in the masked areas, based on the information content of the observed sky regions, being a natural consequence of Wiener filtering. As described in Section \ref{section2.2}, the algorithm generates a fluctuation map to compensate for the suppressed power on the small scales due to the noise and masked regions of the sky. The resulting constrained realization map, after combining the Wiener-filtered and fluctuation maps, has the correct statistical properties consistent with the simulated input map, as illustrated in Fig. \ref{fig:pol_maps}. 

The power spectra of the dual messenger reconstruction for temperature and polarization are provided in the left panel of Fig. \ref{fig:C_l_recon}, thereby showing that the Wiener-filtered maps can be augmented to constrained realizations, resulting in unbiased power spectra. In the high signal-to-noise regime, the power spectrum of the full-sky, noiseless map is determined by the data, while in the low signal-to-noise regime, by the assumed power spectrum. The main issue with the PCG reconstruction, depicted in the right panel, is that it fails to raise the power associated with $\mathcal{B}$ modes, with the $\mathcal{E}$-mode power spectrum also displaying some artefacts, especially on the large and intermediate scales. The low quality of the PCG reconstruction is evident in the behaviour of the convergence diagnostics for the PCG solution as discussed below. 

To illustrate the convergence behaviour of the algorithms, we provide the variations of the residual errors given by $\left \| \myvec{s}_{i+1} - \myvec{s}_{i} \right\| / \left\| \myvec{s}_{i} \right\|$ and $\left \| \mathcal{A} \myvec{x} - \myvec{y} \right\| / \left\| \myvec{y} \right\|$, for a linear system of equations given by $\mathcal{A} \myvec{x} = \myvec{y}$, in Fig. \ref{fig:convergence_diagnostics}. The residual error given by the latter criterion better characterizes the accuracy of the final solution. We find that, for the dual messenger scheme, this residual error always decreases as the iterations proceed, demonstrating the unconditional stability of the algorithm. The oscillations in the residual error provided by the Cauchy criterion, as explained in KLW17, are due to the cooling scheme implemented. The peaks result from the transitions in the systems of equations with the varying covariance of the auxiliary field, with the residual error always dropping sharply after each peak. However, the oscillations in the residual errors for the PCG implementation indicate its vulnerability to instabilities sourced by numerical noise. The residual error in the final PCG solution does not drop below $10^{-3}$, three orders of magnitude higher than that achieved by the dual messenger solution, implying that the accuracy of the PCG solution remains nevertheless significantly inadequate. 

The corresponding variation in the $\chi^2$ is also displayed in the right panel of Fig. \ref{fig:convergence_diagnostics}. The $\chi^2_{\mathrm{DM}}$ of the dual messenger solution drops rapidly in accordance with the cooling scheme implemented and the final value matches $\langle \chi^2_{\mathrm{d.o.f}} \rangle$, the expectation value of the $\chi^2$, given by the number of degrees of freedom (d.o.f), for the final solution. This is not the case, however, for the PCG solution, with $\chi^2_{\mathrm{PCG}}$ failing to attain the expected level. The convergence diagnostics discussed above demonstrate the effectiveness and quality of polarized signal reconstruction via the dual messenger algorithm. 

The solution may be further refined via the modified Jacobi relaxation scheme described in Appendix \ref{jacobi_appendix}. This Jacobi adaptation was found to be stable with $\epsilon = 10^{-5}$ and lower, but the extra computational cost required, for this specific problem, is not justified due to the inherently high quality of the dual messenger solution. This may nevertheless be of general interest for other Wiener filtering applications, to yield adequate solutions for a significantly reduced number of iterations. 

While the choice of another preconditioner may yield an improved solution for the PCG implementation, in practice, it is highly non-trivial to construct an effective preconditioner, especially when dealing with an ill-conditioned system, as considered in this work. Moreover, as illustrated above, the PCG method remains susceptible to numerical instabilities. 

\subsection{Separation of $\mathcal{E}$ and $\mathcal{B}$ modes}
\label{section3.3}

We briefly discuss the so-called $\mathcal{E}$-$\mathcal{B}$ coupling problem, whereby the $\mathcal{E}$-mode power leaks into the much smaller $\mathcal{B}$-mode power, that plagues well-known approximate methods such as the pseudo-$C_{\ell}$ methods \citep[e.g.][]{bond1998estimating}. This aliasing of power is due to the fact that the spherical harmonics are not orthogonal on a sky with masked regions \citep[e.g.][]{zaldarriaga2001nature, bunn2002detectability, bunn2003pixelised}. However, exact methods such as Gibbs sampling circumvent this predicament.  

The standard procedure of obtaining a complete sky sample involves two steps: A Wiener filter is first used to filter out the noise and reconstruct a clean map. Second, the power loss due to noise and incomplete sky coverage is compensated by a random fluctuation term. The combination of the Wiener-filtered and fluctuation maps subsequently yields a full-sky, noiseless sample that is consistent with observations, i.e. constrained realization (cf. Section \ref{section2.2}). The corresponding $\mathcal{E}$ and $\mathcal{B}$ maps resulting from the dual messenger algorithm are depicted in Fig. \ref{fig:EB_maps}. The $\mathcal{E}$-$\mathcal{B}$ coupling issue no longer arises since we now have a full-sky sample. Since the Wiener filter depends on the choice of an input power spectrum, a Gibbs sampling scheme, where samples of power spectrum would be drawn conditional on the data itself, would simultaneously yield the posterior probability distributions of constrained realizations and their power spectra. This Bayesian framework therefore allows for a statistically optimal separation of $\mathcal{E}$ and $\mathcal{B}$ modes in terms of power spectra.

The dual messenger algorithm can thus be incorporated in such a Gibbs sampling scheme, for instance, as described by \cite{larson2007estimation}, for optimal power spectrum inference from high-resolution polarized CMB data sets. \cite{larson2007estimation} implemented a PCG method, while considering uncorrelated noise only, but the preconditioner was limited by the signal-to-noise ratio of the data and was not sufficiently efficient for joint analysis of temperature and polarization data. We have demonstrated that the dual messenger technique is not hindered by such limitations and maintains its efficiency in performing the two key steps described above, even for high-resolution maps, while accounting for correlated noise, as quantitatively substantiated in the next section.

A recent work by \cite{bunn2016pure}, whereby they demonstrate that pure $\mathcal{E}$ and $\mathcal{B}$ maps, free from any cross-contamination, may be obtained via a Wiener filtering approach, provides further motivation for the polarized Wiener filter. This approach provides real-space maps of the $\mathcal{E}$ and $\mathcal{B}$ modes while conventional methods are limited to power spectrum estimation \citep[e.g.][]{challinor2005error, smith2006pure, smith2006pseudo, smith2007general, grain2009polarized} or produce only the derivatives of the polarization maps \citep[e.g.][]{kim2010EB, zhao2010separating, kim2011how, bowyer2011finite}. Other wavelet-based reconstruction methods \citep[e.g.][]{cao2009wavelet, rogers2016spin, leistedt2017wavelet} must be carefully adapted for the specific problem being investigated. The dual messenger algorithm can therefore be optimized to yield pure $\mathcal{E}$ and $\mathcal{B}$ maps via the framework proposed by \cite{bunn2016pure}.

\subsection{Computational performance}
\label{section3.4}

\begin{table}
    \caption{\label{tab:performance_diagnostics_DM_pol} This table provides the computational performance diagnostics for a series of convergence criteria ($\epsilon$).}
    \begin{center}
    \begin{tabular}{lllll}
        \hline
        $\epsilon$ & $i$ & $t (\mathrm{mins})$ & $\chi^2_{\mathrm{final}}$ & $\frac{\left \| \mathcal{A} \myvec{x}_{\mathrm{final}} - \myvec{y} \right\|} {\left\| \myvec{y} \right\|}$ \\
        \hline \hline
        $10^{-4}$ & $304$ & $10$ & $3.1300 \times 10^6$ & $3.1 \times 10^{-5}$ \\
        $10^{-5}$ & $3589$ & $31$ & $3.1207 \times 10^6$ & $3.9 \times 10^{-6}$ \\
        $10^{-6}$ & $49076$ & $309$ & $3.1196 \times 10^6$ & $6.3 \times 10^{-7}$ \\
        \hline
    \end{tabular}
    \end{center}
\end{table}

The execution times and number of iterations required by the dual messenger algorithm, for $N_{\mathrm{pix}}=1024^2$, to run to completion on a single core of a standard workstation, Intel Core i5-4690 CPU (3.50 GHz), for a series of convergence criteria $\epsilon$, are provided in Table \ref{tab:performance_diagnostics_DM_pol}. The $\chi^2$ and residual errors corresponding to the final solutions are also displayed. While the polarization analysis above was carried out with $\epsilon = 10^{-6}$, such a stringent convergence criterion is not required by most Wiener filtering applications, so execution time is drastically reduced. Code parallelization is another key option to speed up the execution for high-resolution data sets. 

It is worth pointing out that even with more lenient criteria, the algorithm provides decent results. For instance, executing the above computation with $\epsilon = 10^{-4}$ requires only $304$ iterations for convergence, corresponding to few minutes of computation time, and already correctly reconstructs the power spectra on all scales, visually indistinguishable from that displayed in the left panel of Fig. \ref{fig:C_l_recon}. The point at which the threshold $| \Delta \chi^2 | \la \sigma_{\chi^2}$ is attained, where $| \Delta \chi^2 | = | \chi^2 - \langle \chi^2_{\mathrm{d.o.f}} \rangle |$ and $\sigma_{\chi^2} \sim \sqrt{2 n_{\mathrm{d.o.f.}}}$ for $n_{\mathrm{d.o.f.}}$ degrees of freedom, is another performance indicator. The $\chi^2$ for the $\epsilon = 10^{-4}$ run drops below this level in $246$ iterations, again highlighting the reliability of the solution achievable with such a low number of iterations. This remarkable performance of the dual messenger technique is especially significant for applications involving exact inference, such as Gibbs sampling.

There is a minor caveat, nonetheless, concerning the current iteration scheme with the relaxed convergence threshold. When dealing with masked regions and very low noise amplitude per pixel, this may result in a marginal lack of convergence on the largest scales for the temperature map, where more iterations are required to improve the reconstruction. The $\mathcal{E}$ and $\mathcal{B}$ maps and their associated power spectra are, however, unaffected by this issue of convergence. We therefore plan to devise an enhanced iteration scheme, such as an adaptive dual messenger algorithm, to improve the treatment of the temperature mask when dealing with real data sets on the sphere.

The algorithmic complexity of the polarized version of the dual messenger method now reflects the fact that the number of Fourier transforms required per iteration is enhanced by a factor of three. The algorithm now requires six Fourier transforms, $\mathcal{O}(N_{\textup{pix}} \log N_{\textup{pix}})$, and two scalar multiplications corresponding to algebraic operations of $\mathcal{O}(3N_{\textup{pix}})$, per iteration. In terms of memory requirements, temporary storage of two vectors of dimension $3N_{\textup{pix}}$ is required. In contrast, the PCG algorithm requires nine Fourier transforms and ten scalar multiplications, per iteration, while eight vectors of size $3N_{\textup{pix}}$ must be temporarily stored in memory.

\section{Conclusions}
\label{section4}

We demonstrate that our recently proposed dual messenger algorithm maintains its efficiency for the data analysis of joint temperature and polarization maps with correlated noise and reduced sky coverage. This preconditioner-free method deals with ill-conditioned systems, typically encountered in the data analysis of polarized CMB maps, with relative ease. This is particularly important as conventional conjugate gradient solvers fail to deal efficiently with this increase in the condition number of the covariance matrices, as illustrated by the failure of the PCG method implemented in this work to converge to a sensible solution. 

The dual messenger algorithm can be conveniently adapted to generate constrained Gaussian realizations of the CMB sky, which is an essential component of present-day CMB data analyses. The reconstruction of $\mathcal{E}$ and $\mathcal{B}$ maps from the resulting full-sky, noiseless sample avoids the leakage issue that arises due to partial sky coverage. The algorithm can also be extended to perform pure $\mathcal{E}/\mathcal{B}$ decomposition via the approach proposed by \cite{bunn2016pure}. 

The implementation of the dual messenger algorithm is straightforward, while being numerically robust and flexible. We also described how this method can be optimized in a modified Jacobi relaxation scheme to further refine the solution. The dual messenger formalism can be naturally augmented via a further level of sophistication to account for more complex and realistic noise models, such as modulated and (spatially) correlated noise, resulting from the scanning strategy of the instrument. We defer the extension of our algorithm to deal with such noise models, which plague state-of-the-art CMB experiments, to a future investigation.

We intend to demonstrate the numerical implementation of the dual messenger method on the sphere and showcase its application on real data sets to further validate this algorithm as a viable signal reconstruction tool for modern CMB data sets. The dual messenger technique can be further refined to reduce execution time. A potentially significant performance upgrade, especially on the sphere, is provided by the hierarchical framework of the dual messenger algorithm: We adapt the working resolution progressively such that the Nyquist frequency is always slightly above the current $\ell_{\mathrm{iter}}$ considered in the algorithm so as to reduce the cost of harmonic transforms. 

We are hopeful that the development of the dual messenger method will ultimately render exact global Bayesian analyses of high-resolution and high-sensitivity CMB observations numerically tractable and computationally efficient. This high-performance algorithm is particularly adapted to cope with the complex numerical challenges posed by modern data sets and is therefore relevant for current and future high-resolution CMB missions such as Planck, South Pole Telescope, Advanced ACTPol, POLARBEAR, QUBIC, Simons Observatory and CMB-S4. Whilst we have demonstrated the application of this algorithm in a cosmological context, it remains nevertheless relevant for spin field reconstruction in a general framework. 
 
\section*{Acknowledgements}

We thank Franz Elsner, Suvodip Mukherjee and Eric Hivon for interesting discussions and/or for their comments on a draft version of the paper. We acknowledge financial support from the ILP LABEX (under reference ANR-10-LABX-63) which is financed by French state funds managed by the ANR within the Investissements d'Avenir programme under reference ANR-11-IDEX-0004-02. GL also acknowledges financial support from ``Programme National de Cosmologie et Galaxies'' (PNCG) of CNRS/INSU, France, and the ANR BIG4, under reference ANR-16-CE23-0002. The Flatiron Institute is supported by the Simons Foundation. 




\bibliographystyle{mnras} 
\bibliography{./compiled_references} 

\begin{thebibliography}{}
\makeatletter
\relax
\def\mn@urlcharsother{\let\do\@makeother \do\$\do\&\do\#\do\^\do\_\do\%\do\~}
\def\mn@doi{\begingroup\mn@urlcharsother \@ifnextchar [ {\mn@doi@}
  {\mn@doi@[]}}
\def\mn@doi@[#1]#2{\def\@tempa{#1}\ifx\@tempa\@empty \href
  {http://dx.doi.org/#2} {doi:#2}\else \href {http://dx.doi.org/#2} {#1}\fi
  \endgroup}
\def\mn@eprint#1#2{\mn@eprint@#1:#2::\@nil}
\def\mn@eprint@arXiv#1{\href {http://arxiv.org/abs/#1} {{\tt arXiv:#1}}}
\def\mn@eprint@dblp#1{\href {http://dblp.uni-trier.de/rec/bibtex/#1.xml}
  {dblp:#1}}
\def\mn@eprint@#1:#2:#3:#4\@nil{\def\@tempa {#1}\def\@tempb {#2}\def\@tempc
  {#3}\ifx \@tempc \@empty \let \@tempc \@tempb \let \@tempb \@tempa \fi \ifx
  \@tempb \@empty \def\@tempb {arXiv}\fi \@ifundefined
  {mn@eprint@\@tempb}{\@tempb:\@tempc}{\expandafter \expandafter \csname
  mn@eprint@\@tempb\endcsname \expandafter{\@tempc}}}

\bibitem[\protect\citeauthoryear{{Alsing}, {Heavens}, {Jaffe}, {Kiessling},
  {Wandelt}  \& {Hoffmann}}{{Alsing} et~al.}{2016}]{alsing2016hierarchical}
{Alsing} J.,  {Heavens} A.,  {Jaffe} A.~H.,  {Kiessling} A.,  {Wandelt} B.,
  {Hoffmann} T.,  2016, \mn@doi [\mnras] {10.1093/mnras/stv2501}, \href
  {http://adsabs.harvard.edu/abs/2016MNRAS.455.4452A} {455, 4452}

\bibitem[\protect\citeauthoryear{{Anderes}, {Wandelt}  \& {Lavaux}}{{Anderes}
  et~al.}{2015}]{anderes2015bayesian}
{Anderes} E.,  {Wandelt} B.~D.,   {Lavaux} G.,  2015, \mn@doi [\apj]
  {10.1088/0004-637X/808/2/152}, \href
  {http://adsabs.harvard.edu/abs/2015ApJ...808..152A} {808, 152}

\bibitem[\protect\citeauthoryear{{Bond}, {Jaffe}  \& {Knox}}{{Bond}
  et~al.}{1998}]{bond1998estimating}
{Bond} J.~R.,  {Jaffe} A.~H.,   {Knox} L.,  1998, \mn@doi [\prd]
  {10.1103/PhysRevD.57.2117}, \href
  {http://adsabs.harvard.edu/abs/1998PhRvD..57.2117B} {57, 2117}

\bibitem[\protect\citeauthoryear{{Bouchet}, {Prunet}  \& {Sethi}}{{Bouchet}
  et~al.}{1999}]{bouchet1999multifrequency}
{Bouchet} F.~R.,  {Prunet} S.,   {Sethi} S.~K.,  1999, \mn@doi [\mnras]
  {10.1046/j.1365-8711.1999.02118.x}, \href
  {http://adsabs.harvard.edu/abs/1999MNRAS.302..663B} {302, 663}

\bibitem[\protect\citeauthoryear{{Bowyer}, {Jaffe}  \& {Novikov}}{{Bowyer}
  et~al.}{2011}]{bowyer2011finite}
{Bowyer} J.,  {Jaffe} A.~H.,   {Novikov} D.~I.,  2011, preprint, \href
  {http://adsabs.harvard.edu/abs/2011arXiv1101.0520B} {} (\mn@eprint {arXiv}
  {1101.0520})

\bibitem[\protect\citeauthoryear{{Bunn}}{{Bunn}}{2002}]{bunn2002detectability}
{Bunn} E.~F.,  2002, \mn@doi [\prd] {10.1103/PhysRevD.65.043003}, \href
  {http://adsabs.harvard.edu/abs/2002PhRvD..65d3003B} {65, 043003}

\bibitem[\protect\citeauthoryear{{Bunn}}{{Bunn}}{2011}]{bunn2011efficient}
{Bunn} E.~F.,  2011, \mn@doi [\prd] {10.1103/PhysRevD.83.083003}, \href
  {http://adsabs.harvard.edu/abs/2011PhRvD..83h3003B} {83, 083003}

\bibitem[\protect\citeauthoryear{{Bunn} \& {Wandelt}}{{Bunn} \&
  {Wandelt}}{2017}]{bunn2016pure}
{Bunn} E.~F.,  {Wandelt} B.,  2017, \mn@doi [\prd]
  {10.1103/PhysRevD.96.043523}, \href
  {http://adsabs.harvard.edu/abs/2017PhRvD..96d3523B} {96, 043523}

\bibitem[\protect\citeauthoryear{{Bunn}, {Fisher}, {Hoffman}, {Lahav}, {Silk}
  \& {Zaroubi}}{{Bunn} et~al.}{1994}]{bunn1994wiener}
{Bunn} E.~F.,  {Fisher} K.~B.,  {Hoffman} Y.,  {Lahav} O.,  {Silk} J.,
  {Zaroubi} S.,  1994, \mn@doi [\apjl] {10.1086/187515}, \href
  {http://adsabs.harvard.edu/abs/1994ApJ...432L..75B} {432, L75}

\bibitem[\protect\citeauthoryear{{Bunn}, {Zaldarriaga}, {Tegmark}  \& {de
  Oliveira-Costa}}{{Bunn} et~al.}{2003}]{bunn2003pixelised}
{Bunn} E.~F.,  {Zaldarriaga} M.,  {Tegmark} M.,   {de Oliveira-Costa} A.,
  2003, \mn@doi [\prd] {10.1103/PhysRevD.67.023501}, \href
  {http://adsabs.harvard.edu/abs/2003PhRvD..67b3501B} {67, 023501}

\bibitem[\protect\citeauthoryear{{Cao} \& {Fang}}{{Cao} \&
  {Fang}}{2009}]{cao2009wavelet}
{Cao} L.,  {Fang} L.-Z.,  2009, \mn@doi [\apj] {10.1088/0004-637X/706/2/1545},
  \href {http://adsabs.harvard.edu/abs/2009ApJ...706.1545C} {706, 1545}

\bibitem[\protect\citeauthoryear{{Challinor} \& {Chon}}{{Challinor} \&
  {Chon}}{2005}]{challinor2005error}
{Challinor} A.,  {Chon} G.,  2005, \mn@doi [\mnras]
  {10.1111/j.1365-2966.2005.09076.x}, \href
  {http://adsabs.harvard.edu/abs/2005MNRAS.360..509C} {360, 509}

\bibitem[\protect\citeauthoryear{{Dunkley} et~al.,}{{Dunkley}
  et~al.}{2009}]{dunkley2009fiveyear}
{Dunkley} J.,  et~al., 2009, \mn@doi [\apjs] {10.1088/0067-0049/180/2/306},
  \href {http://adsabs.harvard.edu/abs/2009ApJS..180..306D} {180, 306}

\bibitem[\protect\citeauthoryear{{Elsner} \& {Wandelt}}{{Elsner} \&
  {Wandelt}}{2009}]{elsner2010improved}
{Elsner} F.,  {Wandelt} B.~D.,  2009, \mn@doi [\apjs]
  {10.1088/0067-0049/184/2/264}, \href
  {http://adsabs.harvard.edu/abs/2009ApJS..184..264E} {184, 264}

\bibitem[\protect\citeauthoryear{{Elsner} \& {Wandelt}}{{Elsner} \&
  {Wandelt}}{2010}]{elsner2010local}
{Elsner} F.,  {Wandelt} B.~D.,  2010, \mn@doi [\apj]
  {10.1088/0004-637X/724/2/1262}, \href
  {http://adsabs.harvard.edu/abs/2010ApJ...724.1262E} {724, 1262}

\bibitem[\protect\citeauthoryear{{Elsner} \& {Wandelt}}{{Elsner} \&
  {Wandelt}}{2012a}]{elsner2012fastcalculation}
{Elsner} F.,  {Wandelt} B.~D.,  2012a, \mn@doi [\aap]
  {10.1051/0004-6361/201218985}, \href
  {http://adsabs.harvard.edu/abs/2012A%26A...540L...6E} {540, L6}

\bibitem[\protect\citeauthoryear{{Elsner} \& {Wandelt}}{{Elsner} \&
  {Wandelt}}{2012b}]{elsner2012likelihood}
{Elsner} F.,  {Wandelt} B.~D.,  2012b, \mn@doi [\aap]
  {10.1051/0004-6361/201219293}, \href
  {http://adsabs.harvard.edu/abs/2012A%26A...542A..60E} {542, A60}

\bibitem[\protect\citeauthoryear{{Elsner} \& {Wandelt}}{{Elsner} \&
  {Wandelt}}{2013}]{EW12}
{Elsner} F.,  {Wandelt} B.~D.,  2013, \mn@doi [\aap]
  {10.1051/0004-6361/201220586}, \href
  {http://cdsads.u-strasbg.fr/abs/2013A%26A...549A.111E} {549, A111}

\bibitem[\protect\citeauthoryear{{Eriksen} et~al.,}{{Eriksen}
  et~al.}{2004}]{eriksen2004power}
{Eriksen} H.~K.,  et~al., 2004, \mn@doi [\apjs] {10.1086/425219}, \href
  {http://adsabs.harvard.edu/abs/2004ApJS..155..227E} {155, 227}

\bibitem[\protect\citeauthoryear{{Galli} et~al.,}{{Galli}
  et~al.}{2014}]{galli2014cmb}
{Galli} S.,  et~al., 2014, \mn@doi [\prd] {10.1103/PhysRevD.90.063504}, \href
  {http://adsabs.harvard.edu/abs/2014PhRvD..90f3504G} {90, 063504}

\bibitem[\protect\citeauthoryear{{Gleser}, {Nusser}  \& {Benson}}{{Gleser}
  et~al.}{2008}]{gleser2008decontamination}
{Gleser} L.,  {Nusser} A.,   {Benson} A.~J.,  2008, \mn@doi [\mnras]
  {10.1111/j.1365-2966.2008.13897.x}, \href
  {http://adsabs.harvard.edu/abs/2008MNRAS.391..383G} {391, 383}

\bibitem[\protect\citeauthoryear{{Grain}, {Tristram}  \& {Stompor}}{{Grain}
  et~al.}{2009}]{grain2009polarized}
{Grain} J.,  {Tristram} M.,   {Stompor} R.,  2009, \mn@doi [\prd]
  {10.1103/PhysRevD.79.123515}, \href
  {http://adsabs.harvard.edu/abs/2009PhRvD..79l3515G} {79, 123515}

\bibitem[\protect\citeauthoryear{{Guzzetti}, {Bartolo}, {Liguori}  \&
  {Matarrese}}{{Guzzetti} et~al.}{2016}]{guzzetti2016gravitational}
{Guzzetti} M.~C.,  {Bartolo} N.,  {Liguori} M.,   {Matarrese} S.,  2016,
  preprint, \href {http://adsabs.harvard.edu/abs/2016arXiv160501615C} {}
  (\mn@eprint {arXiv} {1605.01615})

\bibitem[\protect\citeauthoryear{{Hanson} et~al.,}{{Hanson}
  et~al.}{2013}]{hanson2013detection}
{Hanson} D.,  et~al., 2013, \mn@doi [Physical Review Letters]
  {10.1103/PhysRevLett.111.141301}, \href
  {http://adsabs.harvard.edu/abs/2013PhRvL.111n1301H} {111, 141301}

\bibitem[\protect\citeauthoryear{{Hinshaw} et~al.,}{{Hinshaw}
  et~al.}{2007}]{hinshaw2007threeyear}
{Hinshaw} G.,  et~al., 2007, \mn@doi [\apjs] {10.1086/513698}, \href
  {http://adsabs.harvard.edu/abs/2007ApJS..170..288H} {170, 288}

\bibitem[\protect\citeauthoryear{{Hirata} \& {Seljak}}{{Hirata} \&
  {Seljak}}{2003}]{hirata2003reconstruction}
{Hirata} C.~M.,  {Seljak} U.,  2003, \mn@doi [\prd]
  {10.1103/PhysRevD.68.083002}, \href
  {http://adsabs.harvard.edu/abs/2003PhRvD..68h3002H} {68, 083002}

\bibitem[\protect\citeauthoryear{{Hirata}, {Padmanabhan}, {Seljak}, {Schlegel}
  \& {Brinkmann}}{{Hirata} et~al.}{2004}]{hirata2004cross}
{Hirata} C.~M.,  {Padmanabhan} N.,  {Seljak} U.,  {Schlegel} D.,   {Brinkmann}
  J.,  2004, \mn@doi [\prd] {10.1103/PhysRevD.70.103501}, \href
  {http://adsabs.harvard.edu/abs/2004PhRvD..70j3501H} {70, 103501}

\bibitem[\protect\citeauthoryear{{Hoffman} \& {Ribak}}{{Hoffman} \&
  {Ribak}}{1991}]{hoffman1991constrained}
{Hoffman} Y.,  {Ribak} E.,  1991, \mn@doi [\apjl] {10.1086/186160}, \href
  {http://adsabs.harvard.edu/abs/1991ApJ...380L...5H} {380, L5}

\bibitem[\protect\citeauthoryear{{Hu}}{{Hu}}{2003}]{hu2003cmb}
{Hu} W.,  2003, \mn@doi [Annals of Physics] {10.1016/S0003-4916(02)00022-2},
  \href {http://adsabs.harvard.edu/abs/2003AnPhy.303..203H} {303, 203}

\bibitem[\protect\citeauthoryear{{Hu} \& {Dodelson}}{{Hu} \&
  {Dodelson}}{2002}]{hu2002cmb}
{Hu} W.,  {Dodelson} S.,  2002, \mn@doi [\araa]
  {10.1146/annurev.astro.40.060401.093926}, \href
  {http://adsabs.harvard.edu/abs/2002ARA%26A..40..171H} {40, 171}

\bibitem[\protect\citeauthoryear{{Hu} \& {White}}{{Hu} \&
  {White}}{1997}]{hu1997primer}
{Hu} W.,  {White} M.,  1997, \mn@doi [\na] {10.1016/S1384-1076(97)00022-5},
  \href {http://adsabs.harvard.edu/abs/1997NewA....2..323H} {2, 323}

\bibitem[\protect\citeauthoryear{{Huffenberger}}{{Huffenberger}}{2017}]{huffenberger2017preconditionerfree}
{Huffenberger} K.~M.,  2017, preprint, \href
  {http://adsabs.harvard.edu/abs/2017arXiv170400865H} {} (\mn@eprint {arXiv}
  {1704.00865})

\bibitem[\protect\citeauthoryear{{Huffenberger} \& {N{\ae}ss}}{{Huffenberger}
  \& {N{\ae}ss}}{2017}]{huffenberger2017cosmic}
{Huffenberger} K.~M.,  {N{\ae}ss} S.~K.,  2017, preprint, \href
  {http://adsabs.harvard.edu/abs/2017arXiv170501893H} {} (\mn@eprint {arXiv}
  {1705.01893})

\bibitem[\protect\citeauthoryear{Jacobi}{Jacobi}{1845}]{jacobi1845ueber}
Jacobi C.~G.,  1845, Astronomische Nachrichten, 22, 297

\bibitem[\protect\citeauthoryear{{Jasche} \& {Lavaux}}{{Jasche} \&
  {Lavaux}}{2015}]{jasche2015matrix}
{Jasche} J.,  {Lavaux} G.,  2015, \mn@doi [\mnras] {10.1093/mnras/stu2479},
  \href {http://adsabs.harvard.edu/abs/2015MNRAS.447.1204J} {447, 1204}

\bibitem[\protect\citeauthoryear{{Jasche} \& {Lavaux}}{{Jasche} \&
  {Lavaux}}{2017}]{jasche2017foreground}
{Jasche} J.,  {Lavaux} G.,  2017, \mn@doi [\aap] {10.1051/0004-6361/201730909},
  \href {http://adsabs.harvard.edu/abs/2017A%26A...606A..37J} {606, A37}

\bibitem[\protect\citeauthoryear{{Jasche} \& {Wandelt}}{{Jasche} \&
  {Wandelt}}{2013}]{jasche2013methods}
{Jasche} J.,  {Wandelt} B.~D.,  2013, \mn@doi [\apj]
  {10.1088/0004-637X/779/1/15}, \href
  {http://adsabs.harvard.edu/abs/2013ApJ...779...15J} {779, 15}

\bibitem[\protect\citeauthoryear{{Jasche}, {Kitaura}, {Wandelt}  \&
  {En{\ss}lin}}{{Jasche} et~al.}{2010}]{jasche2010bayesian}
{Jasche} J.,  {Kitaura} F.~S.,  {Wandelt} B.~D.,   {En{\ss}lin} T.~A.,  2010,
  \mn@doi [\mnras] {10.1111/j.1365-2966.2010.16610.x}, \href
  {http://adsabs.harvard.edu/abs/2010MNRAS.406...60J} {406, 60}

\bibitem[\protect\citeauthoryear{{Jewell}, {Levin}  \& {Anderson}}{{Jewell}
  et~al.}{2004}]{jewell2004application}
{Jewell} J.,  {Levin} S.,   {Anderson} C.~H.,  2004, \mn@doi [\apj]
  {10.1086/383515}, \href {http://adsabs.harvard.edu/abs/2004ApJ...609....1J}
  {609, 1}

\bibitem[\protect\citeauthoryear{{Kamionkowski} \& {Kovetz}}{{Kamionkowski} \&
  {Kovetz}}{2016}]{kamionkowski2016quest}
{Kamionkowski} M.,  {Kovetz} E.~D.,  2016, \mn@doi [\araa]
  {10.1146/annurev-astro-081915-023433}, \href
  {http://adsabs.harvard.edu/abs/2016ARA%26A..54..227K} {54, 227}

\bibitem[\protect\citeauthoryear{{Kamionkowski}, {Kosowsky}  \&
  {Stebbins}}{{Kamionkowski} et~al.}{1997a}]{kamionkowski1997statistics}
{Kamionkowski} M.,  {Kosowsky} A.,   {Stebbins} A.,  1997a, \mn@doi [\prd]
  {10.1103/PhysRevD.55.7368}, \href
  {http://adsabs.harvard.edu/abs/1997PhRvD..55.7368K} {55, 7368}

\bibitem[\protect\citeauthoryear{{Kamionkowski}, {Kosowsky}  \&
  {Stebbins}}{{Kamionkowski} et~al.}{1997b}]{kamionkowski1997probe}
{Kamionkowski} M.,  {Kosowsky} A.,   {Stebbins} A.,  1997b, \mn@doi [Physical
  Review Letters] {10.1103/PhysRevLett.78.2058}, \href
  {http://adsabs.harvard.edu/abs/1997PhRvL..78.2058K} {78, 2058}

\bibitem[\protect\citeauthoryear{{Kim}}{{Kim}}{2011}]{kim2011how}
{Kim} J.,  2011, \mn@doi [\aap] {10.1051/0004-6361/201116733}, \href
  {http://adsabs.harvard.edu/abs/2011A%26A...531A..32K} {531, A32}

\bibitem[\protect\citeauthoryear{{Kim} \& {Naselsky}}{{Kim} \&
  {Naselsky}}{2010}]{kim2010EB}
{Kim} J.,  {Naselsky} P.,  2010, \mn@doi [\aap] {10.1051/0004-6361/201014739},
  \href {http://adsabs.harvard.edu/abs/2010A%26A...519A.104K} {519, A104}

\bibitem[\protect\citeauthoryear{{Kodi Ramanah}, {Lavaux}  \& {Wandelt}}{{Kodi
  Ramanah} et~al.}{2017}]{DKR2017reloaded}
{Kodi Ramanah} D.,  {Lavaux} G.,   {Wandelt} B.~D.,  2017, \mn@doi [\mnras]
  {10.1093/mnras/stx527}, \href
  {http://adsabs.harvard.edu/abs/2017MNRAS.468.1782K} {468, 1782}

\bibitem[\protect\citeauthoryear{{Komatsu}, {Spergel}  \& {Wandelt}}{{Komatsu}
  et~al.}{2005}]{komatsu2005measuring}
{Komatsu} E.,  {Spergel} D.~N.,   {Wandelt} B.~D.,  2005, \mn@doi [\apj]
  {10.1086/491724}, \href {http://adsabs.harvard.edu/abs/2005ApJ...634...14K}
  {634, 14}

\bibitem[\protect\citeauthoryear{{Larson}, {Eriksen}, {Wandelt}, {G{\'o}rski},
  {Huey}, {Jewell}  \& {O'Dwyer}}{{Larson} et~al.}{2007}]{larson2007estimation}
{Larson} D.~L.,  {Eriksen} H.~K.,  {Wandelt} B.~D.,  {G{\'o}rski} K.~M.,
  {Huey} G.,  {Jewell} J.~B.,   {O'Dwyer} I.~J.,  2007, \mn@doi [\apj]
  {10.1086/509802}, \href {http://adsabs.harvard.edu/abs/2007ApJ...656..653L}
  {656, 653}

\bibitem[\protect\citeauthoryear{{Lavaux}}{{Lavaux}}{2016}]{lavaux2016virbius}
{Lavaux} G.,  2016, \mn@doi [\mnras] {10.1093/mnras/stv2915}, \href
  {http://adsabs.harvard.edu/abs/2016MNRAS.457..172L} {457, 172}

\bibitem[\protect\citeauthoryear{{Leistedt}, {McEwen}, {B{\"u}ttner}  \&
  {Peiris}}{{Leistedt} et~al.}{2017}]{leistedt2017wavelet}
{Leistedt} B.,  {McEwen} J.~D.,  {B{\"u}ttner} M.,   {Peiris} H.~V.,  2017,
  \mn@doi [\mnras] {10.1093/mnras/stw3176}, \href
  {http://adsabs.harvard.edu/abs/2017MNRAS.466.3728L} {466, 3728}

\bibitem[\protect\citeauthoryear{{Lewis}}{{Lewis}}{2003}]{lewis2003harmonic}
{Lewis} A.,  2003, \mn@doi [\prd] {10.1103/PhysRevD.68.083509}, \href
  {http://adsabs.harvard.edu/abs/2003PhRvD..68h3509L} {68, 083509}

\bibitem[\protect\citeauthoryear{Lewis, Challinor  \& Lasenby}{Lewis
  et~al.}{2000}]{camb1999}
Lewis A.,  Challinor A.,   Lasenby A.,  2000, \mn@doi [Astrophys. J.]
  {10.1086/309179}, 538, 473

\bibitem[\protect\citeauthoryear{{Lewis}, {Challinor}  \& {Turok}}{{Lewis}
  et~al.}{2002}]{lewis2002analysis}
{Lewis} A.,  {Challinor} A.,   {Turok} N.,  2002, \mn@doi [\prd]
  {10.1103/PhysRevD.65.023505}, \href
  {http://adsabs.harvard.edu/abs/2002PhRvD..65b3505L} {65, 023505}

\bibitem[\protect\citeauthoryear{{Manzotti} et~al.,}{{Manzotti}
  et~al.}{2017}]{manzotti2017cmb}
{Manzotti} A.,  et~al., 2017, \mn@doi [\apj] {10.3847/1538-4357/aa82bb}, \href
  {http://adsabs.harvard.edu/abs/2017ApJ...846...45M} {846, 45}

\bibitem[\protect\citeauthoryear{{Millea}, {Anderes}  \& {Wandelt}}{{Millea}
  et~al.}{2017}]{millea2017bayesian}
{Millea} M.,  {Anderes} E.,   {Wandelt} B.~D.,  2017, preprint, \href
  {http://adsabs.harvard.edu/abs/2017arXiv170806753M} {} (\mn@eprint {arXiv}
  {1708.06753})

\bibitem[\protect\citeauthoryear{{O'Dwyer} et~al.,}{{O'Dwyer}
  et~al.}{2004}]{odwyer2004bayesian}
{O'Dwyer} I.~J.,  et~al., 2004, \mn@doi [\apjl] {10.1086/427386}, \href
  {http://adsabs.harvard.edu/abs/2004ApJ...617L..99O} {617, L99}

\bibitem[\protect\citeauthoryear{{Oh}, {Spergel}  \& {Hinshaw}}{{Oh}
  et~al.}{1999}]{oh1999efficient}
{Oh} S.~P.,  {Spergel} D.~N.,   {Hinshaw} G.,  1999, \mn@doi [\apj]
  {10.1086/306629}, \href {http://adsabs.harvard.edu/abs/1999ApJ...510..551O}
  {510, 551}

\bibitem[\protect\citeauthoryear{{Planck Collaboration} et~al.,}{{Planck
  Collaboration} et~al.}{2016}]{13planck2015}
{Planck Collaboration} et~al., 2016, \mn@doi [\aap]
  {10.1051/0004-6361/201525830}, \href
  {http://adsabs.harvard.edu/abs/2016A%26A...594A..13P} {594, A13}

\bibitem[\protect\citeauthoryear{{Rogers}, {Peiris}, {Leistedt}, {McEwen}  \&
  {Pontzen}}{{Rogers} et~al.}{2016}]{rogers2016spin}
{Rogers} K.~K.,  {Peiris} H.~V.,  {Leistedt} B.,  {McEwen} J.~D.,   {Pontzen}
  A.,  2016, \mn@doi [\mnras] {10.1093/mnras/stw2128}, \href
  {http://adsabs.harvard.edu/abs/2016MNRAS.463.2310R} {463, 2310}

\bibitem[\protect\citeauthoryear{Saad}{Saad}{2003}]{saad2003iterative}
Saad Y.,  2003, Iterative methods for sparse linear systems.
Society for Industrial and Applied Mathematics

\bibitem[\protect\citeauthoryear{{Seljak} \& {Hirata}}{{Seljak} \&
  {Hirata}}{2004}]{seljak2004gravitational}
{Seljak} U.,  {Hirata} C.~M.,  2004, \mn@doi [\prd]
  {10.1103/PhysRevD.69.043005}, \href
  {http://adsabs.harvard.edu/abs/2004PhRvD..69d3005S} {69, 043005}

\bibitem[\protect\citeauthoryear{{Seljak} \& {Zaldarriaga}}{{Seljak} \&
  {Zaldarriaga}}{1997}]{seljak1997signature}
{Seljak} U.,  {Zaldarriaga} M.,  1997, \mn@doi [Physical Review Letters]
  {10.1103/PhysRevLett.78.2054}, \href
  {http://adsabs.harvard.edu/abs/1997PhRvL..78.2054S} {78, 2054}

\bibitem[\protect\citeauthoryear{{Seljebotn}, {Mardal}, {Jewell}, {Eriksen}  \&
  {Bull}}{{Seljebotn} et~al.}{2014}]{seljebotn2014multi}
{Seljebotn} D.~S.,  {Mardal} K.-A.,  {Jewell} J.~B.,  {Eriksen} H.~K.,   {Bull}
  P.,  2014, \mn@doi [\apjs] {10.1088/0067-0049/210/2/24}, \href
  {http://adsabs.harvard.edu/abs/2014ApJS..210...24S} {210, 24}

\bibitem[\protect\citeauthoryear{{Seljebotn}, {B{\ae}rland}, {Eriksen},
  {Mardal}  \& {Wehus}}{{Seljebotn} et~al.}{2017}]{seljebotn2017multi}
{Seljebotn} D.~S.,  {B{\ae}rland} T.,  {Eriksen} H.~K.,  {Mardal} K.-A.,
  {Wehus} I.~K.,  2017, preprint, \href
  {http://adsabs.harvard.edu/abs/2017arXiv171000621S} {} (\mn@eprint {arXiv}
  {1710.00621})

\bibitem[\protect\citeauthoryear{{Smith}}{{Smith}}{2006a}]{smith2006pure}
{Smith} K.~M.,  2006a, \mn@doi [\nar] {10.1016/j.newar.2006.09.015}, \href
  {http://adsabs.harvard.edu/abs/2006NewAR..50.1025S} {50, 1025}

\bibitem[\protect\citeauthoryear{{Smith}}{{Smith}}{2006b}]{smith2006pseudo}
{Smith} K.~M.,  2006b, \mn@doi [\prd] {10.1103/PhysRevD.74.083002}, \href
  {http://adsabs.harvard.edu/abs/2006PhRvD..74h3002S} {74, 083002}

\bibitem[\protect\citeauthoryear{{Smith} \& {Zaldarriaga}}{{Smith} \&
  {Zaldarriaga}}{2007}]{smith2007general}
{Smith} K.~M.,  {Zaldarriaga} M.,  2007, \mn@doi [\prd]
  {10.1103/PhysRevD.76.043001}, \href
  {http://adsabs.harvard.edu/abs/2007PhRvD..76d3001S} {76, 043001}

\bibitem[\protect\citeauthoryear{{Smith}, {Zahn}  \& {Dor{\'e}}}{{Smith}
  et~al.}{2007}]{smith2007background}
{Smith} K.~M.,  {Zahn} O.,   {Dor{\'e}} O.,  2007, \mn@doi [\prd]
  {10.1103/PhysRevD.76.043510}, \href
  {http://adsabs.harvard.edu/abs/2007PhRvD..76d3510S} {76, 043510}

\bibitem[\protect\citeauthoryear{{Tegmark}}{{Tegmark}}{1997a}]{tegmark1997power}
{Tegmark} M.,  1997a, \mn@doi [\prd] {10.1103/PhysRevD.55.5895}, \href
  {http://adsabs.harvard.edu/abs/1997PhRvD..55.5895T} {55, 5895}

\bibitem[\protect\citeauthoryear{{Tegmark}}{{Tegmark}}{1997b}]{tegmark1997mapmaking}
{Tegmark} M.,  1997b, \mn@doi [\apjl] {10.1086/310631}, \href
  {http://adsabs.harvard.edu/abs/1997ApJ...480L..87T} {480, L87}

\bibitem[\protect\citeauthoryear{{Wandelt}, {Larson}  \&
  {Lakshminarayanan}}{{Wandelt} et~al.}{2004}]{wandelt2004global}
{Wandelt} B.~D.,  {Larson} D.~L.,   {Lakshminarayanan} A.,  2004, \mn@doi
  [\prd] {10.1103/PhysRevD.70.083511}, \href
  {http://adsabs.harvard.edu/abs/2004PhRvD..70h3511W} {70, 083511}

\bibitem[\protect\citeauthoryear{{Wiener}}{{Wiener}}{1949}]{wiener1949extrapolation}
{Wiener} N.,  1949, Extrapolation, interpolation, and smoothing of stationary
  time series.
Vol. 2. MIT press, Cambridge, MA

\bibitem[\protect\citeauthoryear{{Zaldarriaga}}{{Zaldarriaga}}{2001}]{zaldarriaga2001nature}
{Zaldarriaga} M.,  2001, \mn@doi [\prd] {10.1103/PhysRevD.64.103001}, \href
  {http://adsabs.harvard.edu/abs/2001PhRvD..64j3001Z} {64, 103001}

\bibitem[\protect\citeauthoryear{{Zaldarriaga} \& {Seljak}}{{Zaldarriaga} \&
  {Seljak}}{1997}]{zaldarriaga1997allsky}
{Zaldarriaga} M.,  {Seljak} U.,  1997, \mn@doi [\prd]
  {10.1103/PhysRevD.55.1830}, \href
  {http://adsabs.harvard.edu/abs/1997PhRvD..55.1830Z} {55, 1830}

\bibitem[\protect\citeauthoryear{{Zhao} \& {Baskaran}}{{Zhao} \&
  {Baskaran}}{2010}]{zhao2010separating}
{Zhao} W.,  {Baskaran} D.,  2010, \mn@doi [\prd] {10.1103/PhysRevD.82.023001},
  \href {http://adsabs.harvard.edu/abs/2010PhRvD..82b3001Z} {82, 023001}

\makeatother
\end{thebibliography}



\appendix

\section{Jacobi Relaxation}
\label{jacobi_appendix}

We provide a brief review of the iterative Jacobi technique and describe how it can be implemented with the dual messenger algorithm for a refinement of the solution.

Jacobi relaxation is a well-known iterative technique to solve linear systems of equations \citep{jacobi1845ueber, saad2003iterative}. The conventional Jacobi method cannot be applied to the ill-conditioned system investigated here. We therefore augment the dual messenger algorithm via a modified Jacobi relaxation scheme. 

The standard Jacobi iterative method for the solution at two consecutive iterations, denoted by $n$, is as follows:
\begin{align*}
	\myvec{s}_{\text{\tiny {\textup{WF}}}}^{n+1} &= \myvec{s}_{\text{\tiny {\textup{WF}}}}^n + \Wtilde(\myvec{d} - \mathpzc{W}^{+}\myvec{s}_{\text{\tiny {\textup{WF}}}}^n )  \\
	&= \myvec{s}_{\text{\tiny {\textup{WF}}}}^n + \Wtilde \myvec{d} - \Wtilde \hat{\myvec{d}}, \numberthis
	\label{eq:jacobi_DMPol}
\end{align*}
where the $\mathpzc{W}^{+}$ operator is given by $\mathpzc{W}^{+} = \mymat{N}^{-1}(\mymat{S}^{-1} + \mymat{N}^{-1})$ and $\hat{\myvec{d}} = \mathpzc{W}^{+} \myvec{s}_{\text{\tiny {\textup{WF}}}}^n$. We split the conventional Jacobi operation into two parts; a usual dual messenger ($\Wtilde$) operation on the data $\myvec{d}$ and another $\Wtilde$ operation on the Jacobi correction term. For masked regions, where formally no inverse of $\mymat{N}$ exists, $\mathpzc{W}^{+}$ is the pseudo-inverse, i.e. $\mathpzc{W}\mathpzc{W}^{+} = \Pi$, where $\Pi$ is a projector. 

We first compute the usual dual messenger solution, keeping track of the number of iterations required for each signal covariance truncation $\mu$, then drive the second operation using this prescription for the number of iterations to ensure that the $\Wtilde$ operator is invariant. For numerical stability, we also modify the second equation (\ref{eq:reduced_hybrid_messenger_2nd_equation_DMPol}), after substituting $\myvec{d} \rightarrow \hat{\myvec{d}}$, as follows:
\begin{equation}
	\myvec{t}_{i+1, \myvec{x}} = \mathfrak{D}^{\dagger} \left[\mymat{T} \mathfrak{D} \mathpzc{Z}^n + (\bar{\mathcal{N}}^{-1} + \mymat{T}^{-1}) \mathfrak{D} \mymat{T}^{-1}\myvec{s}_{i+1} \right]_{\myvec{x}}^{-1},
	\label{eq:modified_DM_equation_DMPol}
\end{equation} 
where 
\begin{multline}
	\mathpzc{Z}^n_{\myvec{x}} = {\mathcal{F}^{\dagger}}^{-1} \mathfrak{R}^{\dagger} \mathcal{S}^{-1/2}_{\myvec{\ell}} \left( \mathbb{1} + \mathcal{S}^{1/2}_{\myvec{\ell}} \mathfrak{R} \mathcal{F}^{\dagger} \mathcal{N}^{-1}_{\myvec{x}} \mathcal{F} \mathfrak{R}^{\dagger} \mathcal{S}^{1/2}_{\myvec{\ell}} \right) \\ \cdot \left( \mathcal{S}^{-1/2} \mathfrak{R} \hat{\myvec{s}}_{\text{\tiny {\textup{WF}}}}^n \right)_{\myvec{\ell}}
	\label{eq:jacobi_coeff_DMPol}
\end{multline} 
and including all basis transformations, consistent with the notation employed in Algorithm \ref{alg:dual_messenger_DMPol}. The above rewriting ensures that the equation is free from any singularities.

\section{Incomplete sky coverage}
\label{mask_formalism_appendix}

To account for different temperature and polarization masks, the procedure for solving the messenger equation (\ref{eq:reduced_hybrid_messenger_2nd_equation_DMPol}) must be modified, although the covariance $\mymat{T}$ of the messenger field $\myvec{t}$ is still computed as given in Algorithm \ref{alg:dual_messenger_DMPol}. The noise covariance $\mymat{N}$ can be written as $\mymat{N} =  \bm{\Sigma} \mymat{C}  \bm{\Sigma} = \bm{\Sigma} \mymat{P}^{\dagger} \bm{\Delta} \mymat{P} \bm{\Sigma}$, where $\bm{\Sigma}$ is a diagonal matrix with the eigenvalues $\{ \sigma_{\mathcal{I}}^i, \sigma_{\mathcal{Q}}^i, \sigma_{\mathcal{U}}^i \}$ corresponding to the noise amplitudes for the $i$th pixel, with the orthonormal decomposition of $\mymat{C}$ yielding a diagonal matrix $\bm{\Delta}$. We then obtain $\bar{\mymat{N}} = \mymat{N} - \mymat{T}$ as follows:
\begin{align*}
    \bar{\mymat{N}} &= \bm{\Sigma} \mymat{P}^{\dagger} \bm{\Delta} \mymat{P} \bm{\Sigma} - \mymat{T} \\
	&= \bm{\Sigma} \mymat{P}^{\dagger} ( \bm{\Delta} - \alpha \mymat{P} \bm{\Sigma}^{-2} \mymat{P}^{\dagger} ) \mymat{P} \bm{\Sigma} , \numberthis
	\label{eq:mask_formalism_N_bar_DMPol}
\end{align*}
where, as a reminder, $\mymat{T} = \alpha \mathbb{1}$, with $\alpha$ being the smallest eigenvalue of $\mymat{N}$. To solve the messenger equation (\ref{eq:reduced_hybrid_messenger_2nd_equation_DMPol}), we require the inverse $\bar{\mymat{N}}^{-1}$, 
\begin{equation}
	\bar{\mymat{N}}^{-1} = \bm{\Sigma}^{-1} \mymat{P}^{\dagger} ( \bm{\Delta} - \alpha \mymat{P} \bm{\Sigma}^{-2} \mymat{P}^{\dagger} )^{-1} \mymat{P} \bm{\Sigma}^{-1},
	\label{eq:mask_formalism_inv_N_bar_DMPol}
\end{equation}
such that $\bar{\mymat{N}}^{-1}$ has a block-diagonal structure in pixel space. We obtain the solution for the messenger field by simply evaluating equation (\ref{eq:reduced_hybrid_messenger_2nd_equation_DMPol}) in pixel space,  
\begin{equation}
	\myvec{t}_{\myvec{x}} = \left( \bar{\mymat{N}}^{-1} + \mymat{T}^{-1} \right)_{\myvec{x}}^{-1} \left( \mymat{T}_{\myvec{x}}^{-1} \mathcal{F} \myvec{s}_{\myvec{\ell}} + \bar{\mymat{N}}_{\myvec{x}}^{-1} \myvec{d}_{\myvec{x}} \right) .
	\label{eq:messenger_equation_block_diagonal_DMPol}
\end{equation}

We implement the temperature and polarization masks by increasing the noise variance to infinity for masked pixels, or numerically by setting the inverse noise covariance to zero. This is achieved by setting $\bm{\Sigma}^{-1}|_{\mathrm{mask}} = 0$, subsequently ensuring that data from masked regions do not contaminate the messenger field. 

\section{Preconditioner}
\label{preconditioner_appendix}

For the PCG computation, we implement a straightforward generalization of the diagonal preconditioner presented in Appendix C of KLW17, which also provides a brief review of the PCG method. Essentially, the Wiener filter equation (\ref{eq:wf_equation_DMPol}) is reformulated as $\mathcal{A} \myvec{x} = \myvec{y}$, where $\mathcal{A} = \mathbb{1} + \mymat{S}^{1/2}\mymat{N}^{-1}\mymat{S}^{1/2}$, $\myvec{x} = \mymat{S}^{-1/2}\myvec{s}_{\text{\tiny {\textup{WF}}}}$ and $\mymat{y} = \mymat{S}^{1/2} \mymat{N}^{-1}\myvec{d}$. Solving for $\myvec{x}$ requires the inverse of $\mathcal{A}$, which can be approximated by a preconditioner $\mathcal{M}$, i.e. $\mathcal{M} \approx \mathcal{A}^{-1}$.

Here, the preconditioner $\mathcal{M}$ has the following block-diagonal form, for all multipole moments $\ell$:
\begin{equation}
	\mathcal{M}_{\ell} = \begin{pmatrix}
1 + \psi C_{\ell}^{TT} & 1 + \psi C_{\ell}^{TE} & 0 \\ 
1 + \psi  C_{\ell}^{TE} & 1 & 0 \\ 
0 & 0 & 1 
\end{pmatrix}, 
	\label{eq:preconditioner_DMPol}
\end{equation} 
with $\psi = L^{-4} N_{\mathrm{pix}} \left( \mathcal{F}^{-1} \mymat{N}^{-1}_{\mathcal{I}} \right)_{\myvec{\ell} = 0}$, where $\mymat{N}_{\mathcal{I}}$ is the noise covariance associated with the Stokes parameter $\mathcal{I}$.


\bsp	
\label{lastpage}
\end{document}